\documentclass[a4paper,fleqn]{cas-dc}
\usepackage[authoryear]{natbib}
\usepackage[running]{lineno}
\def\tsc#1{\csdef{#1}{\textsc{\lowercase{#1}}\xspace}}
\tsc{WGM}
\tsc{QE}
\tsc{EP}
\tsc{PMS}
\tsc{BEC}
\tsc{DE}
\begin{document}
\let\WriteBookmarks\relax
\def\floatpagepagefraction{1}
\def\textpagefraction{.001}
\shorttitle{Rayleigh-B{\'e}nard convective instability}
\shortauthors{A. P. Misra et~al.}
\title [mode = title]{ Rayleigh-B{\'e}nard convective motion of stratified fluids in the Earth's troposphere}                      
%\tnotemark[1]
%\tnotetext[1]{This document is the results of the research
   %project funded by the National Science Foundation.}

%\tnotetext[2]{The second title footnote which is a longer text matter
  % to fill through the whole text width and overflow into
   %another line in the footnotes area of the first page.}

%\author[1,3]{CV Radhakrishnan}[type=editor,
%                        auid=000,bioid=1,
%                        prefix=Sir,
%                        role=Researcher,
%                        orcid=0000-0001-7511-2910]
%%%%%%%%%
\author[1]{A.P. Misra}[orcid=0000-0002-6167-8136]
\cormark[1]
%\fnmark[1]
\ead{apmisra@visva-bharati.ac.in; apmisra@gmail.com}
%\ead[url]{www.cvr.cc, cvr@sayahna.org}
%\credit{Conceptualization of this study, Methodology, Software}
\address[1]{ Department of Mathematics, Siksha Bhavana, Visva-Bharati (A Central University), Santiniketan-731 235, India}
\cortext[cor1]{Corresponding author}
%%%
\author[2]{T.D. Kaladze}
%\cormark[2]
%\fnmark[2]{ Deceased.}
\ead{tamaz_kaladze@yahoo.com (Deceased)}
%\ead[url]{www.cvr.cc, cvr@sayahna.org}
%\credit{Conceptualization of this study, Methodology, Software}
\address[2]{I. Vekua Institute of Applied Mathematics and E. Andronikashvili Institute of Physics, Tbilisi State University, Georgia}
\author[2]{D.T. Kaladze}
%\cormark[2]
%\fnmark[2]
\ead{datokala@yahoo.com}
%\ead[url]{www.cvr.cc, cvr@sayahna.org}
%\credit{Conceptualization of this study, Methodology, Software}
%\address[2]{I. Vekua Institute of Applied Mathematics and E. Andronikashvili Institute of Physics, Tbilisi State University, Georgia}
\author[2]{L. Tsamalashvili}
%\cormark[2]
%\fnmark[2]
\ead{luba_tsamal@yahoo.com}
%\ead[url]{www.cvr.cc, cvr@sayahna.org}
%\credit{Conceptualization of this study, Methodology, Software}
%\address[2]{I. Vekua Institute of Applied Mathematics and E. Andronikashvili Institute of Physics, Tbilisi State University, Georgia}

%%%%%%%%%%%%%%%%%%%%%%%%%%%%%%%%%%%%%%%%%%%%%%%%%%%%%%%%%%%%%%%%%%%%%%%%%%%%%%%%%

\begin{abstract}
Recently, Kaladze and Misra [Phys. Scr. 99 (2024) 085013] showed that the tropospheric stratified fluid flows may be unstable by the effects of the negative temperature gradient and the temperature-dependent density inhomogeneity arising from the thermal expansion. They also predicted that the modification in the Brunt-V{\"a}is{\"a}l{\"a} frequency by the density inhomogeneity can lead to  Rayleigh-B{\'e}nard convective instability in the tropospheric unbounded layers. The purpose of the present work is to revisit the Rayleigh-B{\'e}nard convective instability in more detail by considering both unbounded and bounded tropospheric layers. Starting from a set of fluid equations for incompressible neutral fluids with temperature gradients and using the Boussinesq approximation, we derive the general dispersion relations for Rayleigh-B{\'e}nard convective waves in unbounded and bounded tropospheric domains and analyze them with some particular cases both analytically and numerically. 
 We show that the conditions for instability in these two cases significantly differ. Furthermore, we obtain and analyze the critical values of the Raleigh numbers and the expressions for the instability growth rates of thermal waves in the two cases. In the case of the bounded region, we also derive the necessary boundary conditions and note that the vertical wave number is quantified, and the corresponding eigenvalue problem is well-set. 

\end{abstract}
%%%%%%%%%%%%%%%%%%%%%%%%%%
%\begin{highlights}
%\item Convective motion of stratified fluids in the Earth’s troposphere is studied
%\item Instability can occur in both unbounded and bounded tropospheric layers
%\item Critical values of the Raleigh numbers and the instability growth rates are obtained and analysed  
%\item The results should help understand different types of weather systems and their severity in unstable conditions
%\end{highlights}
%%
\begin{keywords}
Convective instability \sep Buoyancy force \sep Stratified fluids \sep Brunt-V{\"a}is{\"a}l{\"a} frequency \sep  Thermal expansion \sep  Earth's troposphere
\end{keywords}
%%%%%% To disable tumbnails
\ExplSyntaxOn
\keys_set:nn { stm / mktitle } { nologo }
\ExplSyntaxOff
%%%%%%
\maketitle
%\date
\section{Introduction} \label{sec-intro}
Convective instability of atmospheric fluids has several key consequences, including affecting the vertical motion of stratified fluids \citep{chitre1973}. As the instability develops, the vertical movements of fluids tend to increase due to thermal expansion, resulting in a large-scale turbulent fluid flow, the formation of extensive vertical clouds, or severe weather conditions such as thunderstorms. In this context, there were several attempts to develop the relevant theories of incompressible atmospheric stratified fluids under gravity to study the instability conditions by the effects of thermal expansion, the density inhomogeneity, the background thermal gradient as well as the interplay between the buoyancy force and the dissipation due to kinematic viscosity and/or thermal diffusivity \citep{zhang2024,shmerlin2013,kameyama2013,kopp2021}. In the subsequent developments of the theory of atmospheric fluid dynamics, Kaladze and Misra \citep{kaladze2023,kaladze2024} showed how the temperature-dependent density inhomogeneity due to thermal expansion and background thermal gradient can significantly modify the well-known Brunt-V{\"a}is{\"a}l{\"a} (BV) frequency, associated with the internal gravity waves, and can play a significant role for the onset of instability of stratified fluids. On the other hand, several authors have also studied and advanced the theory of Darcy-B{\'e}nard convective instability in porous media (See, e.g., \citep{turky2023a,turky2023b,turky2024}     
\par  
Recently, Kaladze and Misra \citep{kaladze2024} studied the influence of temperature-dependent density inhomogeneity due to thermal expansion on the stability of atmospheric stratified fluids owing to its significant applications in the dynamics of atmospheric waves and instability affecting Earth's climate system and showed considerable modification in the Brunt-V{\"a}is{\"a}l{\"a} (BV) frequency. As a result, the necessary conditions for the instability were modified from the previous investigation \citep{kaladze2023}. They also predicted the possibility of the emergence of Rayleigh-B{\'e}nard convective instability of incompressible flows in unbounded tropospheric layers.    
It should be noted that the Rayleigh-B{\'e}nard convection can be represented as a Nambu-metriplectic problem by incorporating dissipative effects \citep{bihlo2008}.   
\par 
The purpose of the present work is to revisit the previous theory \citep{kaladze2024} and advance it by exploring the theory of Rayleigh-B{\'e}nard convective motion and instability in details in both unbounded and bounded domains of the troposphere. We show that the instability conditions and the critical values of the Rayleigh number in unbounded and bounded layers significantly differ. While the vertical wave number can assume any values in an unbounded domain, the same in a bounded domain is an eigenvalue problem. The corresponding instability growth rates are obtained and analyzed. We also show the generation of zero-frequency waves and neutral waves with a finite real wave frequency and zero growth or damping rate.  
%%%%%%%%%%%%%%%%%%%%%%%%%%%%%%%%
\section{Basic equations} \label{sec-basic-eqs}
We consider the convective motion of incompressible stratified neutral gas under gravity in the Earth's troposphere ($0-15$ km) with the effects of the vertical temperature gradient $(d\overline{T}/dz)$ and the temperature-dependent density inhomogeneity due to thermal expansion. Generally, all atmospheric fluid motions are subject to the thermal lapse rate, $\Gamma=-d\overline{T}/dz$. However, we distinguish convective waves from convective instability by assuming a real-wave frequency for the former and a zero real-wave frequency but a finite growth rate to mean the latter.   Our starting point is the following set of fluid equations for the perturbed quantities in the Boussinesq approximation \citep{kaladze2024}.
\begin{equation}\label{eq-moment}
\frac{\partial {\bf v}}{\partial t}+\left({\bf v}\cdot\nabla\right){\bf v}=-\frac{1}{\overline{\rho}}\nabla p^\prime+\beta T^\prime g\hat{z}+\nu\nabla^2{\bf v}, 
\end{equation} 
\begin{equation}\label{eq-cont}
\nabla\cdot{\bf v}=0,
\end{equation} 
\begin{equation}\label{eq-T}
 \frac{\partial T^\prime}{\partial t}+\left({\bf v}\cdot\nabla\right)T^\prime-\Gamma w=\chi\nabla^2T^\prime,
\end{equation} 
where ${\bf v}\equiv(u,v,w)={\bf v}_{\perp}+w\hat{z}$, $\rho^\prime$, $p^\prime$, and $T^\prime$ are, respectively, the perturbed components of the velocity, density, thermal pressure, and temperature of neutral fluids. Also,  
$\beta=- \left({1}/{\overline{\rho}}\right)\left({\partial \rho}/{\partial T}\right)_{p}$ is the coefficient of thermal expansion at a constant pressure, ${\bf g}\equiv g\hat{z}=(0,0,-g)$ is the gravity force per unit fluid mass density acting vertically downward, $\overline{\rho}\equiv \overline{\rho}(z)$ is the inhomogeneous unperturbed or mean fluid density, and $\nu$ is the coefficient of the fluid kinematic viscosity. In Eq. \eqref{eq-moment}, we have used the following equation of state for the temperature dependent perturbed density:   
  \begin{equation}
\rho^\prime({\bf r},t)=-\overline{\rho}(z)\beta T^\prime, \label{eq-state}
\end{equation}
where ${\bf r}=(x,y,z)$, and the term proportional to $\beta$ in Eq. \eqref{eq-moment} represents the buoyancy force, mainly responsible for the convective motion.  Furthermore, in Eq. \eqref{eq-T}, the thermal lapse rate $\Gamma$ can be positive (negative) for a negative (positive) temperature gradient and $\chi$ is the coefficient of the thermal diffusivity.  
For some relevant details, readers are referred to the work of Kaladze and Misra [See Eqs. (12), (13), and (17) of \citep{kaladze2024}]. 
 \par 
 Following Ref. \citep{kaladze2024} and assuming the length scale of variation of $\overline{\rho}(z)$ is much larger than that of $p^\prime$, we obtain the following evolution equation for the perturbed fluid pressure.
  \begin{equation}
\begin{split}
&\left(\frac{\partial}{\partial t}-\chi\Delta\right)\left(\frac{\partial}{\partial t}-\nu\Delta\right)\Delta p^\prime-\beta g\Gamma \Delta_\perp p^\prime=\\
&-\frac{1}{g}\left(\frac{\partial}{\partial t}-\chi\Delta\right)\left(\frac{\partial}{\partial t}-\nu\Delta\right)N^2\frac{\partial p^\prime}{\partial z}, \label{eq-main}
\end{split}
\end{equation} 
where $N^2$ is the Brunt-V{\"a}is{\"a}l{\"a} frequency, given by \citep{kaladze2024},
\begin{equation}\label{eq-N}
\begin{split}
N^2(z)=&-\frac{1}{\overline{\rho}}\frac{d\overline{\rho}}{d z}g,\\
=&-g\left(\frac{1}{\rho_0}\frac{d\rho_0}{d z} -\frac{\beta {\cal T}}{1-\beta {\cal T}} \frac{1}{{\cal T}}\frac{d{\cal T}}{d z}\right).
\end{split}
\end{equation}
Here, $\rho_0\equiv \rho_0(z)$ is the inhomogeneous fluid mass density at $\overline{T}(z)=T_0$ (where $T_0$ can be considered as a reference temperature relevant for internal gravity waves) and  ${\cal T}\equiv{\cal T}(z)=\overline{T}(z)-T_0$. We note that the modification of the BV frequency occurs by the effects of thermal expansion (proportional to $\beta$.) Our aim is not to discuss more about the characteristics of the BV frequency already in Ref. \citep{kaladze2024} but to study its role in the convective wave motion and instability as in the following two subsections \ref{sec-sub-unbd} and \ref{sec-sub-bd} by considering both unbounded and bounded layers in the troposphere.  To this end, we assume that the length scale of fluid density inhomogeneity is approximately constant, and so is the squared frequency $N^2$.
\par 
 We mention that the convective instability in an unbounded domain can manifest in several ways. Although the norm of the perturbations grows in time, the perturbations can decay locally at every point in the unbounded domain, i.e., the growing perturbations are transported or convected towards infinity. Also, in many situations, including experiments and numerical simulations, although bounded domains are more relevant than unbounded ones, from a physical point of view, it is pertinent to investigate the instability conditions and the associated critical values of the Rayleigh number at different atmospheric conditions in an unbounded domain and to distinguish the relevant results with those in a bounded one. 
%%%%%%%
%%%%%%%
\subsection{Convective flow in an unbounded domain} \label{sec-sub-unbd}
  To study the propagation characteristics of the convective wave motion, we assume that the pressure perturbation ($p^\prime$) propagates as a plane wave with the wave vector ${\bf k}=(k_x,k_y,k_z)={\bf k}_\perp+k_z\hat{z}$ and the wave frequency $\omega$. Also, since the perturbations can have an infinite extent, the wave amplitude $(p_0)$  can be taken as constant, i.e., independent of the vertical coordinate $z$. Thus, we consider
\begin{equation}
p^\prime({\bf r},t)=p_0\exp\left(i{\bf k}\cdot{\bf r}-i\omega t\right).\label{eq-wave-form-unbd}
\end{equation}
Although constant amplitude is reasonably a good assumption for an unbounded domain, it can vary with  $z$ in the bounded domain, to be discussed later in Sec. \ref{sec-sub-bd}. 
Substituting Eq. \eqref{eq-wave-form-unbd} into Eq. \eqref{eq-main}, we obtain the following dispersion relation for Rayleigh-B{\'e}nard convective waves \citep{kaladze2024}.
\begin{equation}
\left(k^2-i\frac{N^2}{g}k_z\right)\left[\omega^2+i(\chi+\nu)k^2\omega-\chi\nu k^4\right]+\beta g\Gamma k^2_\perp=0. \label{eq-disp-main}
\end{equation}
In what follows, we note the contributions of different physical parameters and forces. From Eq. \eqref{eq-disp-main}, it is clear that the thermal diffusivity (proportional to $\chi$) and the kinematic viscosity (proportional to $\nu$) effects give rise to wave damping in the absence of the buoyancy force (proportional to $\beta$). On the other hand, without these dissipative effects,  the convective wave motion can be unstable by the influences of the modified BV frequency $N^2$  \citep{kaladze2024}. Before analyzing the general dispersion relation (See Case IV) for the characteristics of convective modes, we first consider some particular cases of interest (See Cases I to III). These will enable us to identify several significant contributions from different physical parameters and to distinguish them with the influence of $N^2$.  
%%%%
%%%%
\subsubsection*{Case I: $\chi=\nu=N^2=0$}
In a most simple situation, i.e., in the absence of the dissipative (thermal diffusivity and viscosity) effects and when the background density $\overline{\rho}$ is independent of the vertical coordinate $z$, i.e., $N^2=0$, the dispersion equation \eqref{eq-disp-main} reduces to

\begin{equation}\label{eq-disp-1a}
\omega^2=-\beta g\Gamma\frac{k_\perp^2}{k^2}.
\end{equation} 
Clearly, the propagating mode with real wave frequency exists when $\Gamma<0$ with positive thermal gradient, i.e., in the stratospheric region \citep{kaladze2023}. We can call such waves without any instability growth rate as convective modes. On the other hand, when $\Gamma>0$, which holds for a negative thermal gradient, such as for stratified fluids in the troposphere (See Table 1 of Ref. \citep{kaladze2023}), the convective motion can be unstable with the instability growth rate, given by,
 \begin{equation}\label{eq-disp-1b}
\gamma=\sqrt{\beta g\Gamma}\frac{k_\perp}{k},
\end{equation} 
where we have considered only the positive sign on the right-hand side because there are no dissipative effects due to the thermal diffusion or kinematic viscosity, so the wave can not be damped. From Eq. \eqref{eq-disp-1b}, it is also evident that the convective instability is mainly caused by the buoyancy force, arising due to the thermal expansion of vertically stratified fluids under gravity with a negative thermal gradient (or, $\Gamma>0$). Furthermore, the instability growth rate typically depends on the vertical wave number $k_z$ (Since $\gamma$ varies inversely with $k_z$ hidden in $k$), i.e., the convective flows become more unstable with an enhanced growth rate as the vertical scale, or the wavelength of perturbations tend to increase.  
%%%%%
%%%%%
\subsubsection*{Case II: $\chi=\nu=0,~N^2\neq0$}
We move on to the case when the background density may not be constant but varies with $z$ such that $N^2\neq0$ and without any dissipative effects. This case is similar to one previously studied by Kaladze and Misra \citep{kaladze2024} [See Eqs. (50) and (51) therein]. However, for limpidness, we repeat it here in our discussion. In this case, the BV frequency will specifically play a key role in determining the instability growth in the presence of the Buoyancy force.   Thus, Eq. \eqref{eq-disp-main} reduces to
\begin{equation}\label{eq-disp2}
\left(k^2-ik_z\frac{N^2}{g}\right)\omega^2+\beta g\Gamma k_\perp^2=0,
\end{equation}
which gives the following expressions for the real $(\omega_r=\Re\omega)$ and the imaginary $(\gamma=\Im\omega)$ parts of $\omega$:
\begin{equation}
\omega_r=\pm\frac{1}{\sqrt{2}}\frac{\sqrt{\beta g\Gamma}k_\perp}{K^2_N}\left(K_N^2-k^2\right)^{1/2}, \label{eq-disp2-real}
\end{equation}
\begin{equation}
\gamma=\mp\frac{1}{\sqrt{2}}\frac{\sqrt{\beta g\Gamma}k_\perp}{K^2_N}\left(K_N^2+k^2\right)^{1/2}, \label{eq-disp2-img}
\end{equation}
provided $\omega_r$ and $\gamma$ satisfy the relation:
\begin{equation} \label{eq-omg-gam}
\omega_r\gamma=-\frac{1}{2}\beta\Gamma N^2\frac{k_\perp^2 k_z}{K_N^4}.
\end{equation}
Here, $K_N=\left(k^4+N^4k_z^2/g^2\right)^{1/4}$. Although the forms of $\gamma$ presented here and in Ref. \citep{kaladze2024} look different, they agree after using Eq. \eqref{eq-omg-gam} and a minor correction with the factor $1/2$, which is missing in Eq. (52) of Ref. \cite{kaladze2024}. We note that Eqs. \eqref{eq-disp2-real} and \eqref{eq-disp2-img} are valid when $\omega_r\gamma<0$, i.e., when $\omega_r$ and $\gamma$ have the opposite signs. Thus, for a positive growth rate, $\gamma>0$ and $\omega_r<0$. The latter may indicate a phase shift in the perturbation or that the phase decreases with time. It is also important to note that the growth rate is independent of the sign of $N^2$, implying that the convective flow without any dissipation in an unbounded domain may become unstable irrespective of the sign of the vertical background density gradient, positive or negative.   
 From Eqs. \eqref{eq-disp2-real} and \eqref{eq-disp2-img}, we find that the perturbations due to convective fluid motion can propagate with a finite real wave frequency (unless at $k_z=0$, where the frequency vanishes) but are unstable having a finite growth rate $\gamma>0$.   Interestingly, such a growth rate of convective waves appears larger than the frequency $\omega_r$ (Compare the square root factors in $\omega_r$ and $\gamma$). The large value of the growth rate can be related to the aperiodic fluid motion and higher values of the Rayleigh number (to be defined shortly), where dominant effects of the buoyancy force over the viscous force come into the picture. For a typical tropospheric model \citep{kaladze2023}, we have $|N^2|\sim10^{-4}-10^{-3}/\rm{s}^2$ and $g\sim10~\rm{m/s^2}$  such that $K_N\approx k$ for $k_z\ll1$. In this situation, $\omega_r\approx0$, and one can have the dominant convective instability.   Thus, in absence of the fluid viscosity and thermal diffusivity, the Brunt-V{\"a}is{\"a}l{\"a} frequency, associated with the length scale of the background density inhomogeneity and modified by the temperature-dependent density inhomogeneity due to thermal expansion, can lead to the Rayleigh-B{\'e}nard convective instability with a finite growth rate $\gamma~(>0)$, but no cut-off at a finite value of $k$. From Eq. \eqref{eq-disp2-img}, we also find that as the vertical wave number $k_z$ becomes smaller or the corresponding vertical wavelengths of perturbations tend to increase, the instability growth rate increases. Such an argument contradicts the instability condition, to be studied in Case III below, by defining the Rayleigh number. Furthermore, the instability growth rate increases with increasing values of the thermal expansion coefficient $\beta$ exceeding a critical value $\beta_c\sim0.004$/K \citep{kaladze2024}, which corresponds to $N^2<0$, i.e., the occurrence of the instability of stratified fluids in the troposphere \citep{kaladze2024,kaladze2023}. Thus, we may conclude that the unstable fluid flow in tropospheric stratified layers associated with internal gravity waves may correspond to convective instability due to the key roles of the buoyancy force in which the growing perturbations (with time) get transported towards infinite boundaries.
%%%
%%%
\subsubsection*{Case III: $\chi\neq0,~\nu\neq0,~N^2=0$}
This case is in contrast to Case II, studied before, in which we mainly focus on the effects of the dissipation due to the kinematic viscosity and the thermal diffusivity. To elucidate it, we consider $\chi\neq0,~\nu\neq0$ but $N^2=0$, i.e., we retain the background density as constant.  In this situation, Eq. \eqref{eq-disp-main} reduces to
\begin{equation}
\omega^2+i\omega k^2(\chi+\nu)-\chi\nu k^4+\beta g\Gamma \frac{k_\perp^2}{k^2}=0, \label{eq-disp3}
\end{equation}
which gives $\omega_r\equiv\Re\omega=0$ and $\gamma\equiv\Im\omega\neq0$, where 
\begin{equation}\label{eq-disp3-img}
\gamma_\mp=-\frac{1}{2}\left[ (\chi+\nu)k^2\pm \sqrt{(\chi-\nu)^2k^4+4\beta g\Gamma \frac{k_\perp^2}{k^2}}  \right]
\end{equation}
From Eq. \eqref{eq-disp3-img}, we find that while the upper sign gives a purely damped mode due to the dissipation $(\gamma\equiv\gamma_-)$, the lower sign can give a buoyancy-force (dominating over the dissipation) driven convective instability with $\gamma\equiv\gamma_+>0$ when the second term (square root term) becomes larger in magnitude than the first term [proportional to $(\chi+\nu)$] in the square brackets, i.e., when 
\begin{equation} \label{eq-inst-cond-disp3}
\frac{k^6}{k_\perp^2}<\frac{\beta g\Gamma}{\chi\nu}.
\end{equation}
Thus, the critical wave number at which the instability growth rate vanishes $(\gamma=0)$ can be obtained via the condition:
\begin{equation}\label{eq-cond-gam01}
\frac{k^6}{k_\perp^2}=\frac{\beta g\Gamma}{\chi\nu}.
\end{equation}
Such a critical wave number typically depends on the ratio of the contributions from the buoyancy and dissipative forces. 
In particular, the convective waves,  which are neither damped nor amplified (i.e., $\gamma=0$), can be called neutral waves.  Thus, it is pertinent to introduce the Rayleigh number, which is the ratio between the contributions from the buoyancy force (providing the destabilizing effect) and the dissipation due to the kinematic viscosity and the thermal diffusivity (providing the stabilizing effect),  defined by,
 \begin{equation} \label{eq-Ra}
 Ra\equiv \frac{\beta g \Gamma H^4}{\chi\nu}=\frac{\beta g \Delta T H^3}{\chi\nu},
 \end{equation}
 where $H$ is the height of the tropospheric fluid layer and $\Gamma=-d\overline{T}/dz\sim\Delta T/H>0$. When the Rayleigh number is below its critical value, the fluid convection does not occur, and the heat is transferred only through thermal conduction. However, the convection starts at the critical value, and the heat is transferred through convection when the Rayleigh number is above the critical value. Rayleigh showed that convective instability can occur when the thermal gradient $\Delta T$ is large enough to exceed $Ra$ a particular value. In order to estimate such a value, we recast Eq. \eqref{eq-inst-cond-disp3}  as \citep{satoh2004}
 \begin{equation} \label{eq-Ra1}
 Ra>\frac{k^6}{k_\perp^2}H^4.
 \end{equation}
We have considered the vertical domain unbounded, and it may be reasonable to consider $H$ as the vertical length scale of perturbations. Thus, with the vertical component of the wave number $k_z$, the expression on the right-hand side of Eq. \eqref{eq-Ra1} assumes the smallest value at $k_\perp=k_z/\sqrt{2}$, and we have the critical (minimum) value of the Rayleigh number, above which the convective instability occurs, as
 \begin{equation} \label{eq-Ra1-cric}
 Ra_{\min}=\left(\frac{k^6}{k_\perp^2}H^4\right)_{ k_\perp={k_z}/{\sqrt{2}}}=\frac{27}{4}k_z^4H^4.
 \end{equation}
We observe that the value of $ Ra_{\min}$ gets significantly reduced in the limit of $k_zH\ll1$, i.e., as the vertical wavelength of perturbation significantly exceeds the vertical length scale. It may be true when the vertical domain is unbounded and the wave number $k_z$ can assume any small values for which $Ra$ becomes smaller with smaller values of $k_z$ relative to the inverse of the vertical scale. In this case, the buoyancy force may not be so strong over the dissipation for which the convective instability may become prominent.   Thus, the convective motions of stratified viscous fluids (with constant background density) under gravity in an unbounded domain become more unstable as the vertical scale $H$ of perturbations deepens compared to the perturbation wavelength. However, we do not have any information about the relation between the components $k_x$ and $k_y$ of the wave number $k$ except a value of $k_\perp=\sqrt{k_x^2+k_y^2}$. Comparing the growth rate [Eq. \eqref{eq-disp3-img}] with that obtained in Case I [Eq. \eqref{eq-disp-1b}], we find that the growth rate gets reduced by the effects of the kinematic viscosity and thermal diffusivity, and the reduction can be significant at higher values of $\chi$ and $\nu$. Also, similar to Case II, the growth rate can be increased with increasing values of $\beta$. Physically, while the viscous force (proportional to $\nu$) and the thermal diffusion (proportional to $\chi$) have stabilizing roles, the buoyancy force (proportional to $\beta$) plays destabilizing roles in the wave motion. So, as the values of $\chi$ and $\nu$ get reduced or a value of $\beta$ increases, the wave tends to become more unstable with a higher growth rate of perturbations.
\par
\begin{figure*}
\centering
\includegraphics[width=5.2in, height=2.8in]{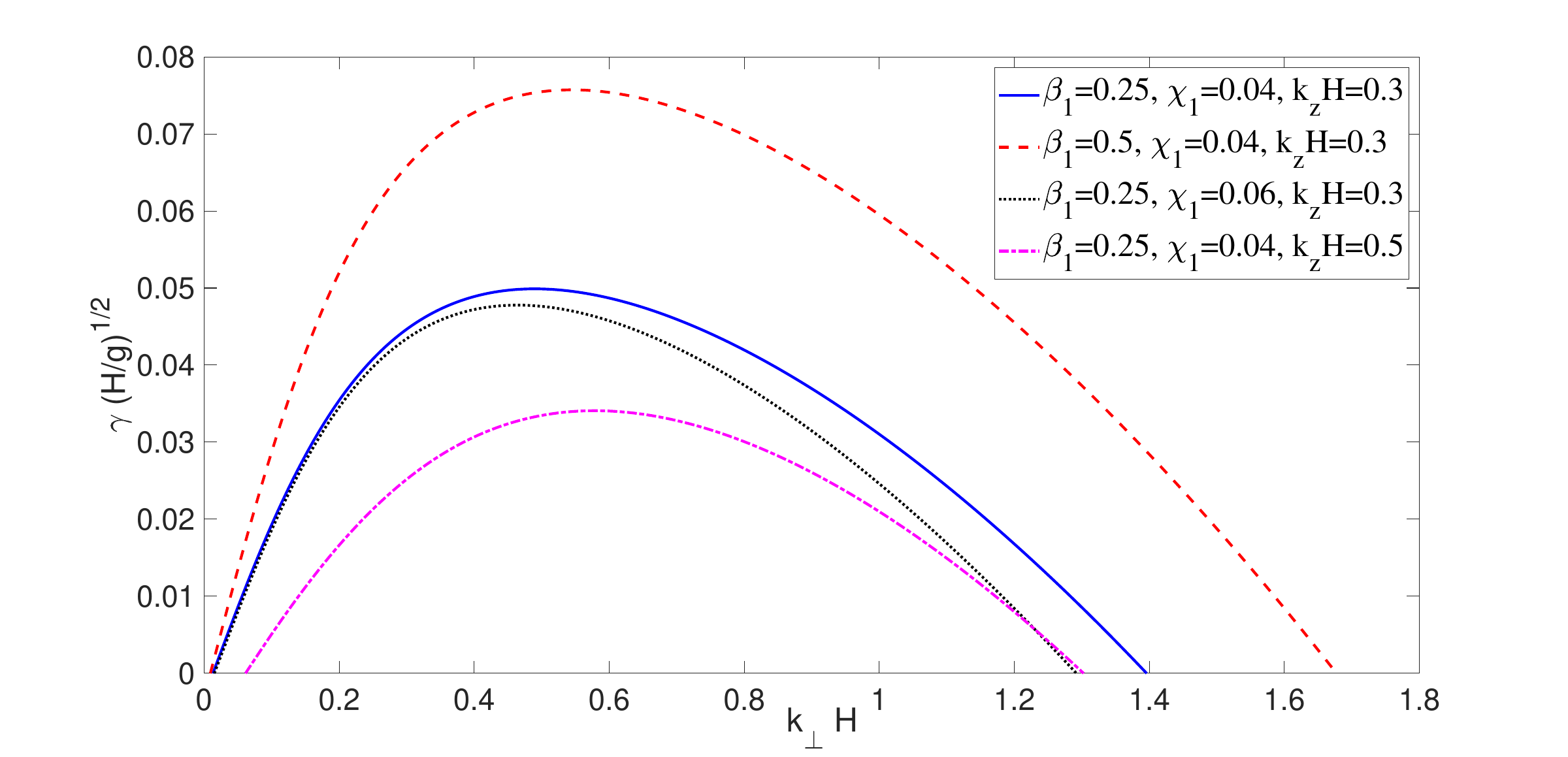}
\caption{The instability growth rate, given by Eq. \eqref{eq-disp3-img} ($\gamma\equiv\gamma_+$), is shown (Case III, Sec. \ref{sec-sub-unbd}: $\chi\neq0,~\nu\neq0,~N^2=0$) for different values of the parameters: $\beta_1\equiv\beta T_0$, $\chi_1\equiv\chi/\sqrt{gH^3}$, and $k_zH$ as in the legend. Here, $H$ is the vertical length scale of perturbation. Note that the parameters $\nu$ and $\Gamma$ have the similar effects on the instability growth rate as of $\chi$ and $\beta$ respectively (not shown). The fixed parameter values are $H=800$ m, $g=9.8~\rm{m/s}^2$, $T_0=250$ K, $\Gamma=0.0065$ K/m, and $\nu_1\equiv\nu/\sqrt{gH^3}=0.03$.   }
\label{fig1}
\end{figure*}

 
For visualization of the characteristics of the growth rate, we have plotted $\gamma$ against the normalized wave number $k_\perp H$ as shown in Fig. \ref{fig1}. It is evident that, in contrast to Cases I and II, where the buoyancy force is the only dominating force, and the perturbations grow with increasing values of $k_\perp H$, the growth rate has cut-offs at smaller and larger wavelengths of perturbations. First of all, from the dependency of $\gamma$ [See Eq. \eqref{eq-disp3-img}] on the parameters, we note that the qualitative behaviors of $\gamma$ by the influences of $\beta$ and $\chi$ are the same as of $\Gamma$ and $\nu$ respectively. Thus, it is sufficient to investigate the qualitative features of $\gamma$ by the effects of the parameters $\beta$, $\chi$, $N^2$ (for the present case, we assume it to be zero), and the vertical wave number $k_zH$. Initially, for fixed values of the parameters $\beta$, $\chi$, and $k_zH$, and a small value of $k_\perp H~(<1)$, the contributions from both the buoyancy and dissipative forces are small, and so is the growth rate (close to zero). As the value of $k_\perp H$ starts increasing, the buoyancy force tends to dominate over the dissipation for which the growth rate gradually increases and reaches a maximum value until the domination is strong. As $k_\perp H$ further increases, the contribution from the dissipation starts increasing, and the growth rate tends to reduce and eventually vanishes at a finite value of $k_\perp H$ (typically $>1$). Thus, we may be conclude that in a dissipative medium of incompressible fluids under gravity in an unbounded domain of the troposphere, the perturbations can grow (leading to the convective instability) only within a finite domain of the wave number $k_\perp H$ due to an interplay between the buoyancy force and the dissipation. On small increasing the value of the thermal expansion coefficient $\beta$, and hence the contribution from the buoyancy force, a significant enhancement of the growth rate having a cut-off at higher $k_\perp H$ is seen (See the dashed line). Thus, as the wave number or the vertical length scale of perturbations deepens, the buoyancy force-driven wave motion tends to become more unstable with higher growth rates. However, such growth rates can get reduced by increasing either $\chi$ (i.e., the enhanced contribution from the dissipation) or the vertical wave number $k_z H$ (Since the contribution of the buoyancy force inversely varies with $k_z H$, it decreases with increasing  $k_z H$; See the dotted and dash-dotted lines).
%%%%%%%%%%%%%%%%%%
%%%%%%%%%%%%%%%%%%%%%%%%%%%%
%%%
%%%%
\subsubsection*{Case IV: $\chi\neq0,~\nu\neq0,~N^2\neq0$}
We turn out to a general situation by taking into account the effects of dissipation due to the kinematic viscosity and the thermal diffusivity as well as the gradient of the background density inhomogeneity, represented by $\chi\neq0,~\nu\neq0,~N^2\neq0$. In this case, the dispersion equation \eqref{eq-disp-main} can be recast as 
\begin{equation}
\omega^2+i(\chi+\nu)k^2\omega-\chi\nu k^4+\frac{\beta g\Gamma k^2_\perp}{K_N^4}\left(k^2+ik_z\frac{N^2}{g}\right)=0. \label{eq-disp4}
\end{equation}
Assuming $\omega=\omega_r+i\gamma$, and separating the real and imaginary parts, we obtain from Eq. \eqref{eq-disp4} the following expressions for $\omega_r$ and $\gamma$.
\begin{equation} \label{eq-disp4-real-img}
\begin{split}
&\omega_r^2=\frac{1}{2}\left[-\Lambda_1+ \sqrt{\Lambda_1^2+\beta^2\Gamma^2N^4\frac{k_z^2 k^4_\perp}{K_N^8}}\right],\\
&\gamma=-\frac{1}{2}(\chi+\nu)k^2\pm\sqrt{\frac{1}{8}(\chi-\nu)^2k^4+\frac{1}{2}\beta g\Gamma \frac{k^2 k^2_\perp}{K_N^4}+\frac{1}{2}\Lambda_2},
 \end{split}
\end{equation}
where $\Lambda_1=({1}/{4})k^4(\chi-\nu)^2+\beta g\Gamma {k^2 k^2_\perp}/{K_N^4}$ and $\Lambda_2=\left(\Lambda_1^2+\beta^2\Gamma^2N^4{k_z^2 k^4_\perp}/{K_N^8}\right)^{1/2}$.
\par
\begin{figure*}
\centering
\includegraphics[width=5.2in, height=2.8in]{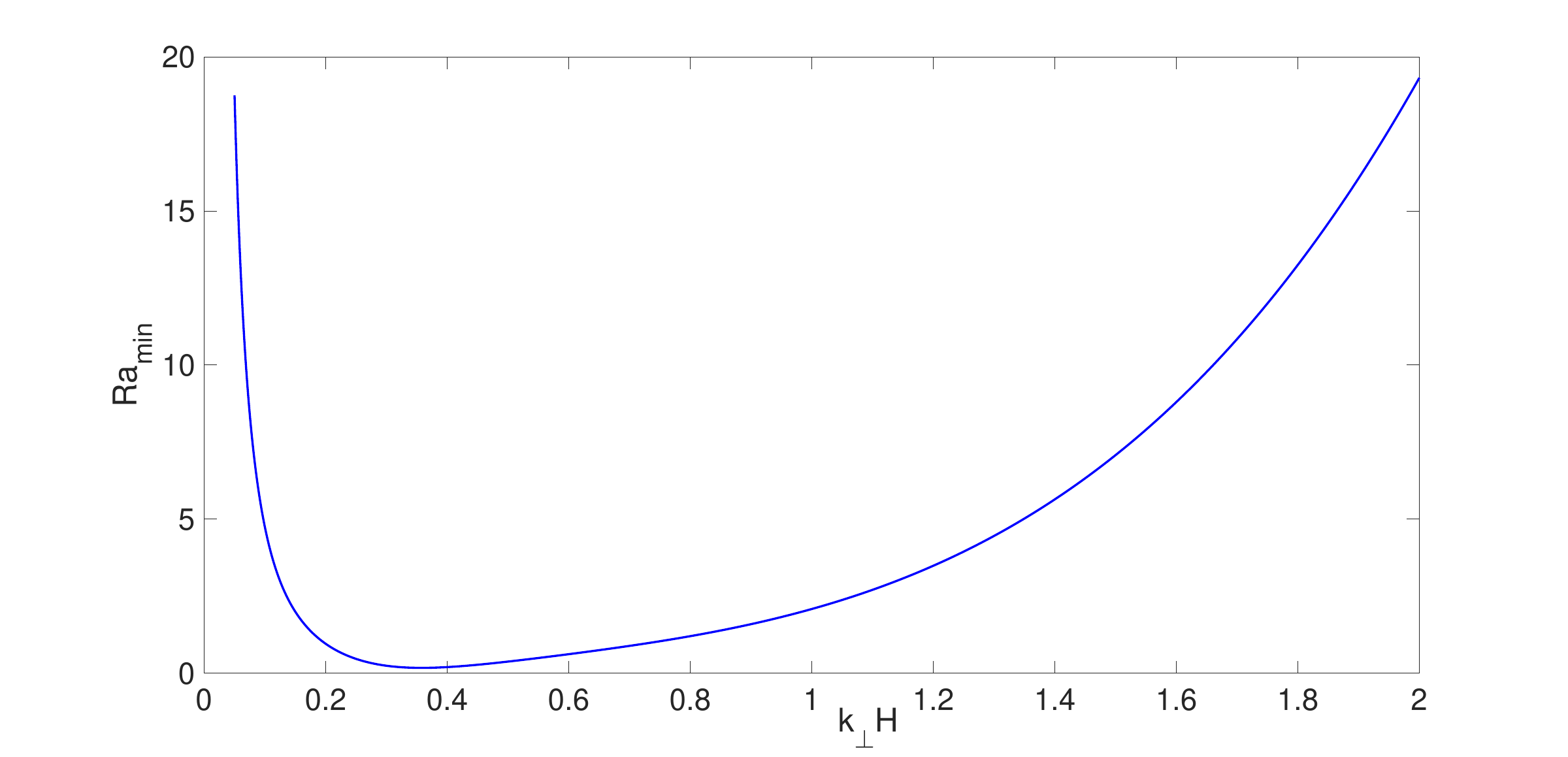}
\caption{The critical value (or minimum value) of the Rayleigh number, given by Eq. \eqref{eq-Ra2-cric}, is plotted against the perpendicular component of the normalized wave number $k_\perp H$. Apart from the sharp peak at a point close to the singular point $k_\perp H=0$, the critical value tends to increase slowly    with increasing values of $k_\perp H$. The fixed parameter values are $\beta=0.001$/K, $H=800$ m, $g=9.8~\rm{m/s}^2$, $T_0=250$ K, $\Gamma=0.0065$ K/m, and $\nu_1\equiv\nu/\sqrt{gH^3}=0.03$, $\chi_1\equiv\chi/\sqrt{gH^3}=0.04$, $k_zH=0.5$, and $N^2=-0.003/\rm{s}^2$. The minimum value of the Rayleigh number is achieved as $Ra_{\rm{min}}\approx0.18$  at $k_\perp H\approx0.35$.}
\label{fig2}
\end{figure*}

 
We note that the lower sign of $\pm$ in $\gamma$ gives a purely damped mode due to the effects of the kinematic viscosity and the thermal diffusivity, which is outside the scope of the present study.   Thus, for the upper sign, the instability condition $(\gamma>0)$ gives 

\begin{equation} \label{eq-Ra2-cond}
\beta g\Gamma \frac{k^2 k^2_\perp}{K_N^4}+\Lambda_2>\frac{1}{4}\left(\chi^2+\nu^2+6\chi\nu\right)k^4,
\end{equation}
which in terms of the Rayleigh number, defined by Eq. \eqref{eq-Ra}, reduces to 
\begin{equation}\label{eq-Ra2-cric}
Ra>Ra_{\rm{min}}\equiv\frac{K_N^4 k^2 H^4}{k_\perp^2}-\frac{\beta^2\Gamma^2N^4k^2_\perp k_z^2H^4}{\chi\nu(\chi+\nu)^2k^6K_N^4}.
\end{equation}
%%%%%%%%%%%%%%%%%%%%%%%%%
From Eq. \eqref{eq-Ra2-cric}, we find that for a given vertical scale height $H$, the term proportional to $\beta$, associated with the buoyancy force due to thermal expansion and background density inhomogeneity, reduces the critical (or minimum) value of the Rayleigh number [\textit{cf}. Eq. \eqref{eq-Ra1}] above which the convective instability can occur. Typically, for $H\sim800$ m, $N^2\sim-0.003/\rm{s}^2$, $g=9.8~\rm{m/s}^2$, $T_0=260$ K, $\Gamma=0.0065$ K/m, $\beta=0.001$/K, $\chi=3000~\rm{m^2/s}$, and $\nu=2000~\rm{m^2/s}$, relevant for tropospheric fluids \citep{kaladze2023}, the variation of the critical values of the Rayleigh number against the normalized wave number $k_\perp H$ are shown in Fig. \ref{fig2}. We see that after the sharp peak at a point close to the singular point $k_\perp H=0$, the critical value of the Rayleigh number decreases, achieving a minimum ($\approx0.18$) at $k_\perp H\approx0.35$, and then tends to increase slowly with increasing values of $k_\perp H$. Thus, given a small but finite value of the wave number $k_\perp$, a larger vertical length scale of perturbations $H$ results in higher values of $Ra_{\rm{min}}$ and the gravitational force to become more dominant over the pressure gradient and viscous forces, which can lead to stronger convective instability. Mathematically, the minimum value of the Rayleigh number can be exceedingly high at  $k_\perp H\lesssim0.1$, i.e., when the wavelength of horizontal perturbations greatly exceeds the vertical scale size $H$. However, this limiting case may not be relevant to unbounded vertical domains (with infinite extent) of the troposphere. We also note that the critical Rayleigh number assumes values less than unity in the interval $0.2\lesssim k_\perp H\lesssim0.7$, corresponding to a regime where the dissipative force is stronger than the buoyancy force. In this case, for the instability to occur, the Rayleigh number can be smaller than unity even when it exceeds the critical value. Such a small value of $Ra<1$ and the corresponding regime of $k_\perp H$ may not be admissible as we require the buoyancy force to be dominant over the dissipative force for stronger instability. Thus, the convective wave motion of tropospheric viscous fluids under gravity becomes more unstable when the wavelengths of horizontal perturbations remain smaller compared to the vertical scale size (i.e., $k_\perp^{-1} <H$).   We note that the critical value of the Rayleigh number does not alter significantly with a small change of the values of the parameters $\beta$ (or $\Gamma$), $\chi$ (or $\nu$), and $N^2$. Also,  at $Ra=Ra_{\rm{min}}$, the growth rate vanishes, and one can have the convective motion of neutral waves for $N^2<0$ with the wave frequency, given by,
\begin{equation}
\omega_r=-\frac{\beta\Gamma k_\perp^2 k_z N^2}{K_N^4k^2(\chi+\nu)}.
\end{equation}
%%%%%%%%%%%%%%%%%%
%%%%%%%%%%%%%%%%%%%%%%%%%%%%
In particular, for $N^2=0$, i.e., when we ignore the density inhomogeneity, the condition \eqref{eq-Ra2-cond} reduces to the same condition as Eq. \eqref{eq-inst-cond-disp3}.
\par
\begin{figure*}
\centering
\includegraphics[width=6.2in, height=2.8in]{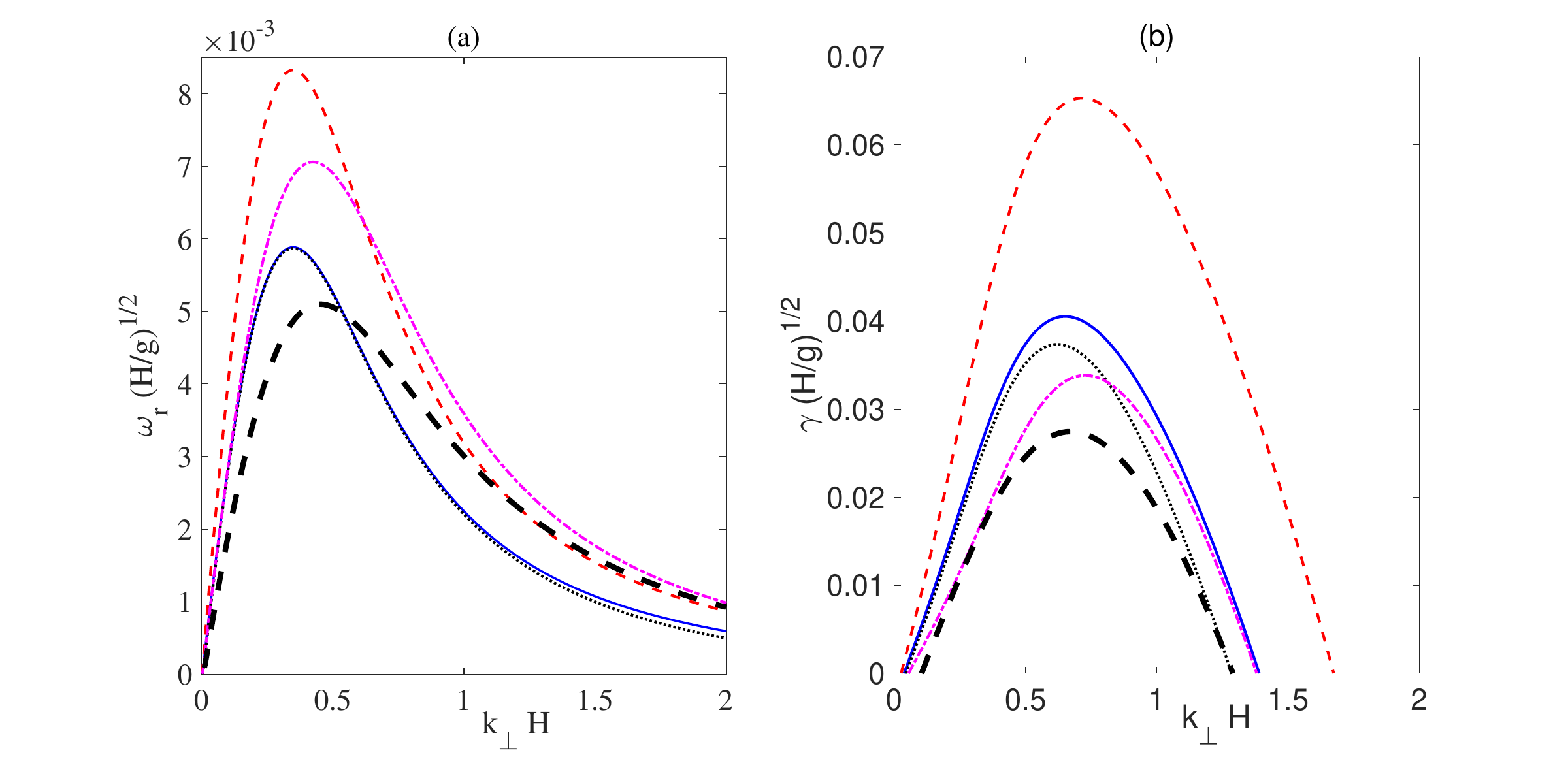}
\caption{The real wave frequency [$\omega_r$, Subplot (a)] and the instability growth rate [$\gamma$, Subplot (b)], given by Eq. \eqref{eq-disp4-real-img}, are shown (Case IV, Sec. \ref{sec-sub-unbd}: $\chi\neq0,~\nu\neq0,~N^2\neq0$) for different values of the parameters: $\beta_1\equiv\beta T_0$, $\chi_1\equiv\chi/\sqrt{gH^3}$, $\omega_g^2=-N^2H/g$, and $k_zH$. The solid, dashed, dotted, dash-dotted, and thick dashed lines, respectively, correspond to the values (i) $\beta_1=0.25,~\chi_1=0.04,~\omega_g^2=0.24$, and $k_zH=0.3$, (ii) $\beta_1=0.5,~\chi_1=0.04,~\omega_g^2=0.24$, and $k_zH=0.3$, (iii) $\beta_1=0.25,~\chi_1=0.06,~\omega_g^2=0.24$, and $k_zH=0.3$, (iv) $\beta_1=0.25,~\chi_1=0.04,~\omega_g^2=0.41$, and $k_zH=0.3$, and (v) $\beta_1=0.25,~\chi_1=0.04,~\omega_g^2=0.24$, and $k_zH=0.5$. Here, $H$ is the vertical length scale of perturbation. Note that the parameters $\nu$ and $\Gamma$ have the similar effects on the instability growth rate as of $\chi$ and $\beta$ respectively (not shown). The fixed parameter values are the same as for Fig. \ref{fig1}, i.e., $H=800$ m, $g=9.8~\rm{m/s}^2$, $T_0=250$ K, $\Gamma=0.0065$ K/m, and $\nu_1\equiv\nu/\sqrt{gH^3}=0.03$.    }
\label{fig3}
\end{figure*}

 
Figure \ref{fig3} shows the profiles of the real wave frequency $\omega_r$ [Subplot (a)]  and the growth rate of instability $\gamma$ [Subplot (b)], given by Eq. \eqref{eq-disp4-real-img}, for different values of the parameters $\beta$, $\chi$, $N^2$, and $k_z$. The influences of the other parameters on the growth rate are not shown for the reason mentioned before in Case III. The authors of Ref. \citep{kaladze2024} discussed the characteristics of  $\omega_r$ and $\gamma$ in the same general case as Case IV. However, they analyzed the results without estimating the critical Rayleigh number above which the convective instability can occur. We somewhat repeat the discussion here in more detail to distinguish from those presented in Cases I-III. So, in contrast to Fig. 3 of Ref. \citep{kaladze2024}, we plot the wave frequency and the growth rate against the normalized horizontal wave number $k_\perp H$ to maintain parity with the previous results in Cases II and III. It is seen from the subplots (a) and (b) of Fig. \ref{fig3} that for each profile, there exists a critical value of $k_\perp H$ below which both the wave frequency $\omega_r$ and the growth rate $\gamma$ increase and attain maximum values (This occurs as the Rayleigh number increases with $k_\perp H$; See Fig. \ref{fig2}) and above which, while $\omega_r$ reaches a steady state value at higher  $k_\perp H$, the growth rate vanishes at a finite value of the same. When the parameter $\beta$ assumes a higher value or the contribution from the buoyancy force remains higher than dissipation, both the wave frequency and the growth rate are significantly enhanced (See the solid and dashed lines). However, when the contribution from the dissipation gets increased by increasing a value of $\chi$, or a value of any one of $N^2$ and $k_z H$ is increased, the growth rate gets notably reduced (Compare the solid line with dotted, dash-dotted, and thick dashed line) except for $\omega_r$, which increases with an increase of the value of $N^2$ [See the solid and dash-dotted lines in Subplot (a)]. We note that the presence of the BV frequency, $N^2$, due to the background density inhomogeneity introduces an additional term in $Ra$, which enhances the critical or minimum value of the Rayleigh number at its increasing values but reduces the growth rate. 
 Similarly, when the vertical wave number gets increased, or the wavelength of perturbation is reduced compared to the vertical scale length, the minimum value of $Ra$ is also increased, but the buoyancy force remains smaller than the dissipative force (\textit{cf}. Fig. \ref{fig2}; $Ra_{\rm{min}}<1$ in $0.2\lesssim k_\perp H\lesssim 0.7$), and eventually resulting in a reduction of the growth rate.  
%%%%%%%%%%%%%%%%%%
%%%%%%%%%%%%%%%%%%%%%%%%%%%%

 \subsection{Convective flow in a bounded domain} \label{sec-sub-bd} 
\begin{figure}
\includegraphics[width=3.2in, height=2.0in]{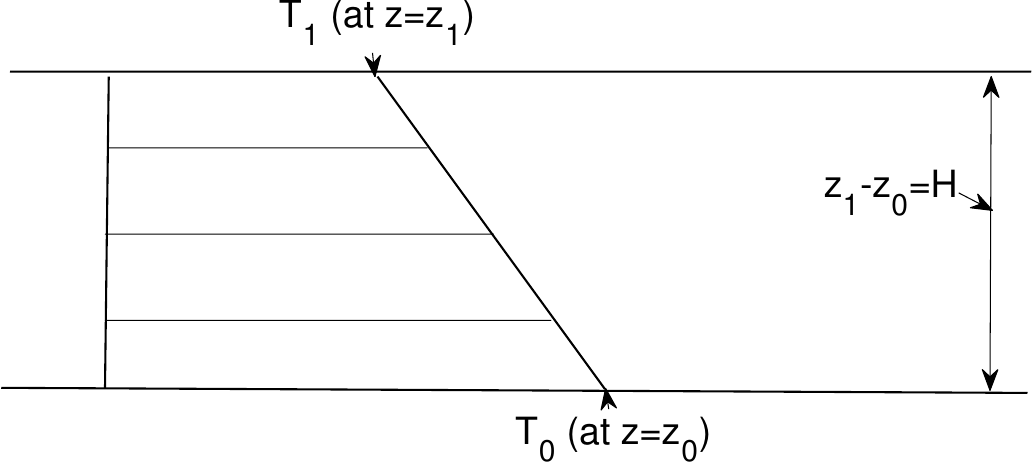}
\caption{A diagram showing the trpospheric bounded region between two walls at $z=z_0$ (with temperature $T_0$) and $z=z_1$ (with temperature $T_1$) such that $z_1-z_0=H$. The temperature sharply decreases with the atmospheric height $z$.}
\label{fig4}
\end{figure}

 It is to be noted that the atmospheric temperature changes sharply with height, i.e., it decreases as the atmospheric height increases from a layer at $z=z_0$ to another layer at $z=z_1$. So, one can consider the tropospheric region to be bounded between two ``walls" at heights $z=z_0$ and $z=z_1$ so that $z_1-z_0=H$ represents the height of the troposphere (See Fig. \ref{fig4}). We assume the background (or mean) temperature, $\overline{T}=T_0$ at $z_0=0$ km  and $\overline{T}=T_1$ at $z_1=15$ km for the tropospheric layers. Furthermore, from the unperturbed state of the heat-energy equation, i.e., $d^2\overline{T}/dz^2=0$, we have \citep{kaladze2024}
\begin{equation} \label{eq-Tz1}
\overline{T}(z)=C_0z+C_1,
\end{equation}
where $C_0$ and $C_1$ are constants such that $C_0=d\overline{T}/dz=-\Gamma$. Here, we repeat, $\Gamma$ is called the lapse rate, and $\Gamma>0$ holds for a negative temperature gradient as relevant for the troposphere \citep{kaladze2023}. Since we have assumed that $\overline{T}=T_0$ at $z_0=0$ km, we define $C_1=T_0$.  Also, from the condition $\overline{T}=T_1$ at $z=z_1=H$ (See Fig. \ref{fig4}), we get $T_1=C_0z_1+C_1$, which gives
 \begin{equation} \label{eq-Gamma}
 \Gamma=\frac{T_0-T_1}{H}.
 \end{equation}
Thus, Eq. \eqref{eq-Tz1} reduces to
\begin{equation} \label{eq-Tz2}
{\cal T}(z)\equiv\overline{T}(z)-T_0=-\Gamma z,
\end{equation}
i.e., the unperturbed temperature inhomogeneity relative to some reference temperature $T_0$ in a bounded domain is directly proportional to the atmospheric height $z$. It follows that the values of $N^2$ will differ in unbounded and bounded regions of atmospheric fluids. 
%%%%%%%%%%%%%%%%%%%%%%%%%%%%%%%%%%%
%%%%%%%%%%%%%%%%%%%%%%%%%%%%%%%%%%%
\par 
Typically, the profiles of vertical perturbations depend on the boundary conditions on the upper and lower ``rigid" surfaces, which are the no-slip conditions, and that the walls are maintained at constant temperatures. Thus, we choose the boundary conditions as 
\begin{equation}
\frac{\partial u}{\partial z}=\frac{\partial v}{\partial z}=0, ~w=0.
\end{equation}
However, the rigid lid conditions $u=v=w=0$ may be more appropriate for practical purpose.
\par 
In what follows, we seek a plane wave solution of Eq. \eqref{eq-main} in the following form:  
\begin{equation}
p^\prime({\bf r},t)=\tilde{p}(z)\exp\left(ik_xx+ik_yy-i\omega t\right), \label{eq-waveform-bdd}
\end{equation}
where, in contrast to the unbounded region (Sec. \ref{sec-sub-unbd}), we have assumed the wave amplitude of perturbations to vary with the $z$-coordinate and the wave propagates in the $xy$-plane. Substitution of Eq. \eqref{eq-waveform-bdd} into Eq. \eqref{eq-main} results in the following reduced equation.
\begin{equation} \label{eq-main-bdd}
\begin{split}
&\left\lbrace\left[\chi\left(\frac{d^2}{dz^2}-k^2_\perp \right)+i\omega\right]\left[\nu\left(\frac{d^2}{dz^2}-k^2_\perp \right)+i\omega\right]\left(\frac{d^2}{dz^2}-k^2_\perp \right)\right.\\
&\left.+\beta g\Gamma k_\perp^2 \right\rbrace\tilde{p}(z)=
-\frac{1}{g}\left[\chi\left(\frac{d^2}{dz^2}-k^2_\perp \right)+i\omega\right]\\
&\times \left[\nu\left(\frac{d^2}{dz^2}-k^2_\perp \right)+i\omega\right]N^2\frac{d\tilde{p}}{dz}.
\end{split}
\end{equation}
For the existence of zero-frequency waves with $\omega=0$ as in Case III of Sec. \ref{sec-sub-unbd}, Eq. \eqref{eq-main-bdd} gives for $N^2=0$ the following equation \citep{satoh2004}.
\begin{equation} \label{eq-main-neutral}
\left[\left(\frac{d^2}{dz^2}-k^2_\perp \right)^3+Ra\frac{k^2_\perp}{H^4}\right]\tilde{p}=0.
\end{equation}
%%%%%%%%%%%%%%%%%%%%%
We also require the linearized forms of Eqs. \eqref{eq-moment}-\eqref{eq-T}, given by,

\begin{equation}
\left(\frac{\partial}{\partial t}-\nu\nabla^2\right)u=-\frac{1}{\overline{\rho}}\frac{\partial p^\prime}{\partial x},\label{eq-u1}
\end{equation}
\begin{equation}
\left(\frac{\partial }{\partial t}-\nu\nabla^2\right)v=-\frac{1}{\overline{\rho}}\frac{\partial p^\prime}{\partial y},\label{eq-v1}
\end{equation}
\begin{equation}
\frac{\partial w}{\partial t}=-\frac{1}{\overline{\rho}}\frac{\partial p^\prime}{\partial z}+\beta g T^\prime+\nu\nabla^2w,\label{eq-w1}
\end{equation}
\begin{equation}
\frac{\partial u}{\partial x}+\frac{\partial v}{\partial y}+\frac{\partial w}{\partial z}=0,\label{eq-cont1}
\end{equation}
\begin{equation}
\frac{\partial T^\prime}{\partial t}-\Gamma w=\chi\nabla^2T^\prime, \label{eq-T1}
\end{equation}
where we have obtained Eqs. \eqref{eq-u1}-\eqref{eq-w1} from Eq. \eqref{eq-moment}  after separating the velocity components along the axes.  Next, substituting Eq. \eqref{eq-waveform-bdd} into Eqs. \eqref{eq-u1}-\eqref{eq-T1} and setting $T^\prime=0$, we obtain the following boundary conditions for $\tilde{p}$. 
\begin{equation}\label{eq-bcs}
\frac{d\tilde{p}}{dz}=\frac{d^3\tilde{p}}{dz^3}=\frac{d^5\tilde{p}}{dz^5}=\cdots=0 ~\rm{at}~z=0,H.
\end{equation}
A plane wave solution of  Eq. \eqref{eq-bcs} for $\tilde{p}$  is given by
\begin{equation} \label{eq-p-sol-bc}
\tilde{p}(z)=\tilde{p}_0\cos{(mz)};~m=\frac{(2n+1)\pi}{H},~ n=0,\pm1,\pm2,...,
\end{equation}
where $\tilde{p}_0=\tilde{p}(0)$ is a constant and $m$ stands for the vertical wave number. Here, we choose a different symbol for the vertical wave number to distinguish from that for the unbounded region. 
%which is written as a linear combination of Eq. \eqref{eq-wave-form-unbd}. 
Substituting the solution form of perturbation \eqref{eq-waveform-bdd} into Eq. \eqref{eq-main-neutral}, we obtain the following relation between the Rayleigh number and the wave number $k_\perp$ \citep{satoh2004}.
\begin{equation} \label{eq-Ra-bdd}
Ra=\frac{(k_\perp^2+m^2)^3}{k_\perp^2}H^4.
\end{equation}
For a fixed vertical wave number $m$, $Ra$ takes its minimum value $Ra_{\rm{min}}=(27/4)m^4H^4$ [\textit{cf}. Eq. \eqref{eq-Ra1-cric}] at 
$k_\perp=m/\sqrt{2}$. From Eq. \eqref{eq-p-sol-bc}, $m$ takes the smallest value, $|m|=\pi/H$ at $n=0$ or $-1$. In this case, the critical values of the Rayleigh number and the horizontal wave number, above and below which the convective instability occurs, are given by \citep{satoh2004}
\begin{equation} \label{eq-Rac-kc}
Ra_{\min}=\frac{27}{4}\pi^4\approx 657.5,~~k_{\perp c}=\frac{\pi}{\sqrt{2}H}\approx 2.2/H.
\end{equation}
It follows that in the case of a bounded region, we require a higher value of the Rayleigh number for the instability to occur compared to the bounded region, and the growth rate can achieve a maximum value within the domain $0<k_\perp H\lesssim2.2$. 
Similar to cases III and IV of Sec. \ref{sec-sub-unbd} with $\omega_r=0$ but $\gamma\neq0$, we can obtain expressions for the instability growth rates of convective motions in a bounded region of the troposphere.  From Eqs. \eqref{eq-waveform-bdd} and \eqref{eq-p-sol-bc}, we have
\begin{equation} \label{eq-p-prime}
p^\prime=\tilde{p}_0\cos{(mz)}e^{(ik_x x+ik_y y)} e^{\gamma t}.
\end{equation}
Also, from Eqs. \eqref{eq-u1} and \eqref{eq-v1}, we have the following solutions for the transverse velocity components \citep{satoh2004}.
\begin{equation}
\begin{split}\label{eq-u2v2}
&u=-i\frac{k_x}{\nu k^2+\gamma}\frac{\tilde{p}_0}{\overline{\rho}}\cos{(mz)}e^{(ik_x x+ik_y y)} e^{\gamma t},\\
&v=-i\frac{k_y}{\nu k^2+\gamma}\frac{\tilde{p}_0}{\overline{\rho}}\cos{(mz)}e^{(ik_x x+ik_y y)} e^{\gamma t}.
\end{split}
\end{equation}
Next, to find a solution for the vertical component $w$, we first obtain a reduced equation for $w$. To this end,  we eliminate $T^\prime$ from Eqs. \eqref{eq-w1} and \eqref{eq-T1} to get
\begin{equation}
\begin{split}
&\left[\left(\frac{\partial}{\partial t}-\chi\Delta\right)\left(\frac{\partial}{\partial t}-\nu\Delta\right)-\beta g\Gamma\right]w=\\
&-\left(\frac{\partial}{\partial t}-\chi\Delta\right)\frac{1}{\overline{\rho}}\frac{\partial p^\prime}{\partial z}. \label{eq-w2}
\end{split}
\end{equation}
Thus, using Eq. \eqref{eq-p-prime} for  $p^\prime$, we have from Eq. \eqref{eq-w2} the following solution for $w$.
 \begin{equation} \label{eq-w3}
 \begin{split}
w=&\frac{\gamma+\chi k^2}{\left(\gamma+\chi k^2\right)\left(\gamma+\nu k^2\right)-\beta g \Gamma} \\
&\times m  \frac{\tilde{p}_0}{\overline{\rho}}\sin{(mz)}e^{(ik_x x+ik_y y)} e^{\gamma t}.
\end{split}
\end{equation}
Now, from the dispersion equation \eqref{eq-disp3} of Case III, Sec. \ref{sec-sub-unbd}, the condition for $\omega_r=0$ gives
\begin{equation}
\beta g \Gamma\frac{k^2_\perp}{k^2}=\left(\gamma+\nu k^2\right)\left(\gamma+\chi k^2\right),
\end{equation}
for which Eq. \eqref{eq-w3} reduces to \citep{satoh2004}
\begin{equation} \label{eq-w4}
w=-\frac{k_\perp^2}{\left(\gamma+\nu k^2\right)} 
\frac1m  \frac{\tilde{p}_0}{\overline{\rho}}\sin{(mz)}e^{(ik_x x+ik_y y)} e^{\gamma t}.
\end{equation}
Next, we obtain a solution for $T^\prime$ from Eqs. \eqref{eq-T1} and \eqref{eq-p-prime} as \citep{satoh2004}
\begin{equation} \label{eq-T2}
 \begin{split}
T^\prime=&-\frac{k_\perp^2\Gamma}{m\left(\gamma+\chi k^2\right)\left(\gamma+\nu k^2\right)} \\
&\times \frac{\tilde{p}_0}{\overline{\rho}}\sin{(mz)}e^{(ik_x x+ik_y y)} e^{\gamma t}\\
&\equiv-\widetilde{T}_0\sin{(mz)}e^{(ik_x x+ik_y y)} e^{\gamma t},
\end{split}
\end{equation}
where $\widetilde{T}_0$ represents the wave amplitude of temperature perturbation.  We note that in all the solutions 
\eqref{eq-u2v2}, \eqref{eq-w3}, \eqref{eq-w4}, and \eqref{eq-T2}, $k^2=k^2_\perp+m^2$, where $m$ is given by Eq. \eqref{eq-p-sol-bc}.
\par 
From Eqs. \eqref{eq-w4} and \eqref{eq-T2}, we obtain the following relation between $w$ and $T^\prime$.
\begin{equation}
w=\frac{\gamma+\chi k^2}{\Gamma} T^\prime,
\end{equation} 
which shows that the background thermal gradient and the thermal fluctuation influence the vertical movements of tropospheric fluids in a bounded domain, which is also responsible for the instability to grow instead of damping due to the thermal diffusivity.   
Thus, it is pertinent to define the horizontal average of vertical heat flux as \citep{satoh2004}
\begin{equation}\label{eq-average1}
\overline{(\Re w)(\Re T^\prime)}=\frac{\gamma+\chi k^2}{2\Gamma}\widetilde{T}_0^2\sin^2(mz)e^{2\gamma t}.
\end{equation}
For the critical mode (neutral wave) with $\gamma=0$ [See Eq. \eqref{eq-cond-gam01}] and $k^2=3m^2/2=3\pi^2/2H^2$, Eq. \eqref{eq-average1} gives \citep{satoh2004}
\begin{equation}\label{eq-average2}
\overline{(\Re w)(\Re T^\prime)}=\frac{3\pi^2}{4}\frac{\chi \widetilde{T}_0^2}{\Gamma H^2}\sin^2(mz).
\end{equation}
Next, we introduce the Nusselt number, which is the ratio between the total heat flux and the heat flux due to the thermal conductivity. Since the horizontal average of the conductivity heat flux is $\chi\Gamma$, the Nusselt number for the critical mode is given by \citep{satoh2004}
\begin{equation} \label{eq-Nu}
Nu=\frac{\chi\Gamma+\overline{(\Re w )(\Re T^\prime)}}{\chi\Gamma}=1+\frac{3\pi^2}{4}\frac{\widetilde{T}_0^2}{\Gamma^2 H^2}\sin^2(mz).
\end{equation} 
The amplitude $\widetilde{T}_0$ in the linear regime is small. However, in the nonlinear regime, the amplitude of temperature perturbation is bounded by half of the temperature difference between the top and bottom boundaries, i.e., $\widetilde{T}_0\leq \Gamma H/2$. If we choose the maximum value of the temperature, $\widetilde{T}_{0\max}= \Gamma H/2$, the Nusselt number estimated in Eq. \eqref{eq-Nu} also takes the maximum value, i.e.,  $Nu_{\max}=2.85$, which characterizes a kind of slug flow or laminar fluid flow. In the following, we will consider two different cases separately, similar to Cases III and IV of Sec. \ref{sec-sub-unbd} for the unbounded region.
\subsection*{Case I: $\chi\neq0,~\nu\neq0,~N^2=0$} \label{sec-sub-bdd-caseI}
By disregarding the effects of $N^2$ or when the background fluid density is held constant, Eq. \eqref{eq-main-bdd} reduces to
\begin{equation} \label{eq-bdd-case1}
\begin{split}
&\left\lbrace\left[\chi\left(\frac{d^2}{dz^2}-k^2_\perp \right)+i\omega\right]\left[\nu\left(\frac{d^2}{dz^2}-k^2_\perp \right)+i\omega\right]\right.\\
&\left.\times \left(\frac{d^2}{dz^2}-k^2_\perp \right)+\beta g\Gamma k_\perp^2 \right\rbrace\tilde{p}(z)=0.
\end{split}
\end{equation}
Next, assuming the solution of Eq. \eqref{eq-bdd-case1} to be of the form \eqref{eq-p-sol-bc} and that $\omega=\omega_r+i\gamma$, we get
\begin{equation} \label{eq-omeg-gam-bdd}
\begin{split}
&\omega_r^2-\gamma^2-\gamma(m^2+k^2_\perp)(\chi+\nu)-\chi\nu(m^2+k^2_\perp)^2\\
&+\frac{\beta g\Gamma k^2_\perp}{m^2+k^2_\perp}=0,\\
&\omega_r\left[2\gamma+(m^2+k^2_\perp)(\chi+\nu)\right]=0.
\end{split}
\end{equation}
Since we are looking for a positive growth rate $(\gamma>0)$, the second equation of Eq. \eqref{eq-omeg-gam-bdd} gives $\omega_r=0$. So, from the first equation, we have the following expression for $\gamma$.
\begin{equation}\label{eq-gam-bdd1}
\begin{split}
\gamma=&-\frac{1}{2}(m^2+k^2_\perp)(\chi+\nu)\\
&+ \frac{1}{2}\sqrt{(m^2+k^2_\perp)^2(\chi-\nu)^2+\frac{4\beta g \Gamma k^2_\perp}{m^2+k^2_\perp}}.
\end{split}
\end{equation}
Comparing this expression with that in the unbounded case [Case III, Eq. \eqref{eq-disp3-img} for $\gamma>0$], we find that the wave number $k$ is being replaced by $m^2+k^2_\perp$, i.e., the vertical wave number $k_z$, which assumes any values in the unbounded domains (See Case III of Sec. \ref{sec-sub-unbd}), is replaced by the discrete eigenvalue $m$, and estimating the instability growth rate becomes an eigenvalue problem. Similar to Eq. \eqref{eq-disp3-img}, the minimum or critical value of the Rayleigh number above which the convective instability can occur  is given by [See also Eq. \eqref{eq-Ra-bdd}]
\begin{equation} \label{eq-Ra1-cric}
 Ra_{\min}=\left[\frac{\left(m^2+k^2_\perp\right)^3}{k_\perp^2}H^4\right]_{ k_\perp={m}/{\sqrt{2}}}=\frac{27}{4}m^4H^4,
 \end{equation}
 where $m=(2n+1)\pi/H$ with $n=0,\pm1,\pm2,...$ and the smallest critical value attained at the vertical wave number $mH=\pm\pi$ (corresponding to  $n=0$ or $n=-1$) is approximately $657.5$ [See also Eq. \eqref{eq-Rac-kc}]. It is important to note that, in contrast to the case of the unbounded region, the wave number $mH$ does not assume arbitrary values but discrete eigenvalues corresponding to the integer $n$. The qualitative features of the instability growth rate will remain the same as Fig. \ref{fig1}, and we do not repeat them here for brevity. 
\subsection*{Case II: $\chi\neq0,~\nu\neq0,~N^2\neq0$} \label{sec-sub-bdd-caseII}
This case is similar to Case IV of Sec. \ref{sec-sub-unbd} in which the effects of the BV frequency and dissipation are retained, however, in a bounded region. Substitution of Eq. \eqref{eq-p-sol-bc} into Eq. \eqref{eq-main-bdd} gives
\begin{equation} \label{eq-bdd-case2-main}
\begin{split}
&\left[\chi\left(m^2+k^2_\perp\right)-i\omega\right]\left[\nu\left(m^2+k^2_\perp\right)-i\omega\right]\\
&\times\left(m^2+k^2_\perp\right)-\beta g \Gamma k^2_\perp \\
&= -\frac{N^2}{g}\left[\chi\left(m^2+k^2_\perp\right)-i\omega\right]\left[\nu\left(m^2+k^2_\perp\right)-i\omega\right]\\
&\times m\tan (mz).
\end{split}
\end{equation}
Separating the real and imaginary parts of Eq.  \eqref{eq-bdd-case2-main}, we get 
\begin{equation} \label{eq-bdd-omeg-img}
\begin{split}
&\left(m^2+k^2_\perp\right)\left[\chi\nu\left(m^2+k^2_\perp\right)^2+\gamma\left(\chi+\nu\right)\left(m^2+k^2_\perp\right)\right.\\
&\left.-\omega_r^2+\gamma^2\right]-\beta g\Gamma k^2_\perp+\frac{N^2}{g}m\tan(mz)\\
&\times\left[\chi\nu\left(m^2+k^2_\perp\right)^2+\gamma(\chi+\nu)\left(m^2+k^2_\perp\right)-\omega_r^2+\gamma^2\right]=0,\\
&\omega_r\left\lbrace\left(\chi+\nu\right)\left(m^2+k^2_\perp\right)^2+2\gamma\left(m^2+k^2_\perp\right)\right.\\
&\left.+\frac{N^2}{g}m \tan(mz) \left[\left(\chi+\nu\right)\left(m^2+k^2_\perp\right)+2\gamma\right]\right\rbrace=0.
\end{split}
\end{equation}
If $\omega_r=0$ is a solution of the second equation of Eq. \eqref{eq-bdd-omeg-img}, the first equation gives

\begin{equation} \label{eq-bdd-img1}
\begin{split}
&\left(m^2+k^2_\perp\right)\left[\chi\nu\left(m^2+k^2_\perp\right)^2+\gamma\left(\chi+\nu\right)\left(m^2+k^2_\perp\right)+\gamma^2\right]\\
&-\beta g\Gamma k^2_\perp+\frac{N^2}{g}m\tan(mz)\\
&\times\left[\chi\nu\left(m^2+k^2_\perp\right)^2+\gamma(\chi+\nu)\left(m^2+k^2_\perp\right)+\gamma^2\right]=0, 
\end{split}
\end{equation}
which has a solution for $\gamma$, given by,
\begin{equation} \label{eq-bdd-img2}
\begin{split}
 \gamma&=-\frac12\left(\chi+\nu\right)\left(m^2+k^2_\perp\right)\\
 &+\frac12 \sqrt{\left(\chi-\nu\right)^2\left(m^2+k^2_\perp\right)^2+\frac{4\beta g\Gamma k^2_\perp}{m^2+k^2_\perp+\widetilde{N}^2}}, 
\end{split}
\end{equation}
where we have considered the plus sign before the square root to obtain the growth rate and $\widetilde{N}^2=\left({N^2}/{g}\right)m \tan(mz)$. Thus, the condition for $\gamma\geq0$ gives
\begin{equation} \label{eq-bdd-inst-cond1}
\frac{\beta g\Gamma k^2_\perp}{m^2+k^2_\perp+\widetilde{N}^2}\geq\chi\nu\left(m^2+k^2_\perp\right)^2,
\end{equation}
where the equality sign holds for the critical mode with $\gamma=0$. In terms of the Rayleigh number defined before [Eq. \eqref{eq-Ra}], Eq. \eqref{eq-bdd-inst-cond1} gives for $\gamma>0$ the following condition.
\begin{equation}\label{eq-bdd-inst-cond2}
Ra>\frac{\left(m^2+k^2_\perp\right)^2}{k^2_\perp}H^4\left(m^2+k^2_\perp+\widetilde{N}^2 \right).
\end{equation}
The expression on the right-hand side of Eq. \eqref{eq-bdd-inst-cond2} has a minimum at 
\begin{equation} \label{eq-bdd-k-min}
k^2_\perp=-\frac14\left(m^2+\widetilde{N}^2\right)+\frac14 \sqrt{ \left(5m^2+\widetilde{N}^2\right)^2-16m^4},
\end{equation}
where we have considered the plus sign before the square root, and since $m\tan(mz)=-\left[1/\tilde{p}(z)\right]\left[d\tilde{p}(z)/dz\right]=1/ L_{\tilde{p}}>0$ for $d\tilde{p}(z)/dz<0$ (Typically, the atmospheric pressure decreases with increasing the altitude and  $L_{\tilde{p}}$ denotes the vertical length scale of pressure inhomogeneity) and $N^2<0$ (which holds for the instability of stratified fluids \citep{kaladze2023,kaladze2024}), we have   $\widetilde{N}^2>0$. Thus,   $k^2_\perp>0$ and the right-hand side of Eq. \eqref{eq-bdd-inst-cond2} is also positive.   
  Inserting Eq. \eqref{eq-bdd-k-min} in the right-hand side of   Eq. \eqref{eq-bdd-inst-cond2}, we find
\begin{equation} \label{eq-Ra-min2}
\begin{split}
Ra_{\min}(m)&=\frac{1}{8}H^4\left[ \left(9m^2+\frac{N^2}{g L_{\tilde{p}}} \right){\cal R}+27m^4 \right.\\
&\left.+18\frac{N^2m^2}{g L_{\tilde{p}}}-\frac{N^4}{g^2L^2_{\tilde{p}}}\right],
\end{split}
\end{equation}
where ${\cal R}=\sqrt{\left(5m^2+{N^2}/{g L_{\tilde{p}}} \right)^2-16m^4}$.  It follows that the minimum value of the Rayleigh number depends not only on the BV frequency $N^2$ and the vertical wave number $m$ (eigenvalue) but also on the vertical length scale of pressure inhomogeneity  $L_{\tilde{p}}$.
\par
\begin{figure*}
\centering
\includegraphics[width=5.2in, height=2.8in]{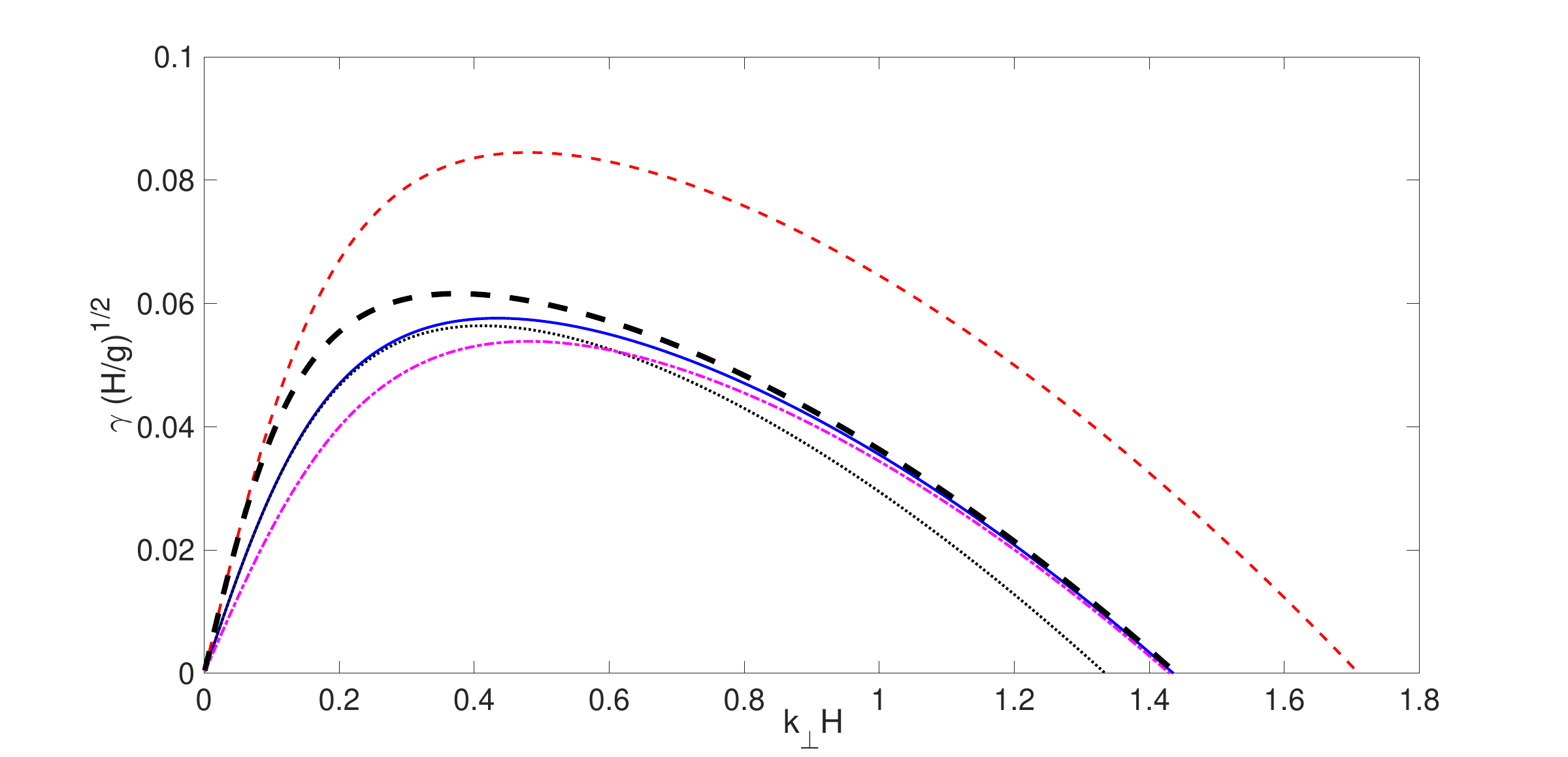}
\caption{The instability growth rate [$\gamma$, given by Eq. \eqref{eq-bdd-img2},is shown (Case II, Sec. \ref{sec-sub-bdd-caseII}): $\chi\neq0,~\nu\neq0,~N^2\neq0$) for different values of the parameters: $\beta_1\equiv\beta T_0$, $\chi_1\equiv\chi/\sqrt{gH^3}$, $\omega_g^2=-N^2H/g$, and $L_{\tilde{p}}$. The solid, dashed, dotted, dash-dotted, and thick dashed lines, respectively, correspond to the values (i) $\beta_1=0.25,~\chi_1=0.04,~\omega_g^2=0.24$, and $k_zH=0.3$, (ii) $\beta_1=0.5,~\chi_1=0.04,~\omega_g^2=0.24$, and $L_{\tilde{p}}=5$, (iii) $\beta_1=0.25,~\chi_1=0.06,~\omega_g^2=0.24$, and $L_{\tilde{p}}/H=5$, (iv) $\beta_1=0.25,~\chi_1=0.04,~\omega_g^2=0.41$, and $L_{\tilde{p}}=5$, and (v) $\beta_1=0.25,~\chi_1=0.04,~\omega_g^2=0.24$, and $L_{\tilde{p}}/H=10$. Here, $H$ is the vertical length scale of perturbation and $L_{\tilde{p}}$ the vertical length scale of pressure inhomogeneity. Note that the parameters $\nu$ and $\Gamma$ have the similar effects on the instability growth rate as of $\chi$ and $\beta$ respectively (not shown). The fixed parameter values are the same as for Fig. \ref{fig1}, i.e., $H=800$ m, $g=9.8~\rm{m/s}^2$, $T_0=250$ K, $\Gamma=0.0065$ K/m, and $\nu_1\equiv\nu/\sqrt{gH^3}=0.03$ together with $m=\pi/H$ at $n=0$.     The critical values corresponding to the solid, dashed, dotted lines are the same, i.e., $Ra_{\min}=659.7$, and to the dash-dotted and thick dashed lines, respectively, are $Ra_{\min}=661.1$ and $658.6$. For $m=3\pi/H$ at $n=1$, and other parameter values as for the solid line, we have $Ra_{\min}=5.3\times10^4$.     }
\label{fig5}
\end{figure*}

 
Figure \ref{fig5} shows the profiles of the instability growth rate $\gamma$, given by Eq. \eqref{eq-bdd-img2}, for different values of the dimensionless parameters: $\beta_1\equiv\beta T_0$, $\chi_1\equiv\chi/\sqrt{gH^3}$, $\omega_g^2=-N^2H/g$, and $L_{\tilde{p}}/H$ (when $d\tilde{p}(z)/dz<0$) that are relevant for tropospheric fluids \citep{kaladze2023} as mentioned in the figure caption. For a fixed set of parameter values, the qualitative feature of the growth rate is similar to Fig. \ref{fig3}(b). We note that the growth rate is significantly enhanced by the effects of increasing values of the thermal expansion parameter $\beta$ and the vertical length scale of pressure inhomogeneity $L_{\tilde{p}}$ compared to that of perturbations $H$ (Compare the solid line with dashed and thick dashed lines). Thus, in contrast to the unbounded domain, the background pressure inhomogeneity effect favors the convective instability with higher growth rates as the corresponding scale size tends to increase compared to the vertical scale of perturbations. Furthermore, the thermal expansion gives rise to a higher growth rate with cut-offs at higher wave numbers relative to the unbounded domain. While the effect of the BV frequency is similar to the bounded domain in reducing the growth rate with its increasing values, the dissipation due to the kinematic viscosity and thermal diffusivity does not have any significant effect in increasing or decreasing the maximum growth rate but giving cut-offs at lower wave numbers as the value of $\chi$ increases (analogous to the case of the bounded domain). We do not discuss the effects of the other parameters on the growth rate for the reason either mentioned in the figure caption or the discussion of Figs. \ref{fig1} and \ref{fig3}.         
%%%%%%%%%%%%%%%%%%
%%%%%%%%%%%%%%%%%%%%%%%%%%%%
\section{Summary and Conclusion}
We have studied the Rayleigh-B{\'e}nard convective wave motion and instability of stratified fluids under gravity in the troposphere as motivated by the recent work of Kaladze and Misra \citep{kaladze2024} in which they showed that the effects of temperature-dependent density inhomogeneity due to thermal expansion can significantly modify the Brunt-V{\"a}is{\"a}l{\"a} (BV) frequency and lead to the onset of convective instability with negative thermal gradient. However, they restricted the model to an unbounded domain of the troposphere and did not explore or discuss many relevant details.  We have revisited the theory of convective motion and instability in more detail by considering both unbounded and bounded tropospheric domains to show that the conditions for instability in these two cases significantly differ. Thus, the previous theory of convective instability in an unbounded domain is generalized and advanced. The critical values of the Raleigh numbers and the expressions for instability growth rates of thermal waves are obtained and analyzed in various parameter regimes of dissipation and thermal expansion associated with the BV frequency in both unbounded and bounded domains. In the latter, we get the necessary boundary conditions and show the quantification of the vertical wave number and the corresponding eigenvalue problem. The main findings can be summarized as follows:
\subsection*{Unbounded domain}
In an unbounded domain of the troposphere, we have considered some particular cases together with a general one to reveal the role of the modified Brunt-V{\"a}is{\"a}l{\"a} frequency on the convective motion and instability.
\begin{itemize}
\item [(i)]  In the simple case, when $\chi=\nu=N^2=0$, convective waves without any damping or instability can propagate only in the stratosphere when $\Gamma<0$, i.e., with a positive thermal gradient. However, in the troposphere with $\Gamma<0$, convective instability can occur with a zero real wave frequency but a finite growth rate. The instability becomes stronger, and the growth rate becomes higher (without any cut-off) as the vertical scale $H$ or the wavelengths of perturbations gradually increase, or the thermal expansion coefficient $\beta$ tends to increase. 
\item [(ii)] When $\chi=\nu=0$ but $N^2\neq0$, the propagation of convective waves with a larger growth rate than the real frequency is possible (i.e., $\gamma>\omega_r$), and this growth rate can be associated with the aperiodic fluid motion and higher values of the Rayleigh number, where dominant effects of the buoyancy force over dissipation come into the picture. Like item (i), the instability growth rate increases with increasing values of $H$ or $\beta$.
\item [(iii)] In contrast to the case of item (ii), i.e., $\chi\neq0,~\nu\neq0$ but $N^2=0$, the convective instability with a zero real frequency but a finite growth rate (i.e., $\omega_r=0,~\gamma>0$)  can occur. The existence of zero-frequency waves with $\omega_r=\gamma=0$ is also possible at some critical value of the wave number,  i.e., they appear as a flat line and may exist everywhere instantly. Furthermore, a minimum, or a critical value of the Rayleigh number, at which the convection starts and above which the instability sets in, is obtained and shown to be proportional to the vertical wave number $k_zH$. We note that as the vertical length scale of perturbations $H$ deepen, or the value of $\chi$ or $\nu$ gets reduced, or that of $\beta$ increases, the convective wave tends to become more unstable with a higher growth rate.
\item [(iv)] In a general situation when $\chi\neq0,~\nu\neq0$, and $N^2\neq0$, convective wave motion and instability are possible by the influence of the BV frequency with $\omega_r,~\gamma>0$. The critical or minimum value of the Rayleigh number gradually increases with the horizontal wave number $k_\perp H$, and as its value increases, the growth rate tends to vanish after achieving a maximum value, implying that the larger the minimum value of the Rayleigh number, the less unstable is the wave motion. 
\end{itemize}
\subsection*{Bounded domain}
In contrast to the unbounded domain, the wave amplitude of perturbation is no longer a constant. Still, it depends on the vertical coordinate and wave number, which assumes discrete eigenvalues. It is seen that the background thermal gradient and the thermal fluctuation influence the vertical movements of tropospheric fluids, which is also responsible for the instability of growing instead of damping due to the thermal diffusivity. The maximum value of the corresponding Nusselt number is approximately $3$, indicating that a kind of slug flow or laminar fluid flow may be possible.  
\begin{itemize}
\item [(i)] In the case of $\chi\neq0,~\nu\neq0$ but $N^2=0$, we have the possibility of the convective instability with $\gamma>0$ and $\omega_r=0$, similar to item (iii) above for an unbounded domain. However, in contrast to (iii), the vertical wave number no longer assumes any values but some discrete eigenvalues. The minimum value of the corresponding Rayleigh number is approximately $658$, which is much higher than predicted in an unbounded domain.  
\item [(ii)]  When $\chi\neq0,~\nu\neq0$, and $N^2\neq0$, we also have the possibility of convective instability with $\gamma>0$ and $\omega_r=0$. However, in addition to the BV frequency $N^2$ and the vertical wave number $m$ (eigenvalue), a new vertical length scale of pressure inhomogeneity $(L_{\tilde{p}})$ appears, which significantly influences the minimum value of the Rayleigh number and the instability growth rate. We have noted that the background pressure inhomogeneity effect favors the convective instability with higher growth rates as the length scale $L_{\tilde{p}}$ tends to increase compared to the vertical scale of perturbations $(H)$.
\end{itemize}
\par 
%To conclude, effects of convective instability in dry tropospheric layers may have consequences including the formation of an inferior mirages and dust and steam devils. 
 To conclude, The present investigation is an attempt to successively improve the existing theory of  wave motion and instability of atmospheric stratified fluids under gravity. In this way, it advances the previous works  \citep{kaladze2023,kaladze2024}. The wave motions and instabilities reported here could help better understand different types of weather systems and their severity in unstable conditions compared to the previous studies. The authors look for further advancement in a medium of rotating fluids \citep{chatterjee2021,abdikian2022,turi2022}.
 %%            $$$$$$$$$$$$$$$$$$$$$$$$$$$$$$$$$$$$$$$$$$$$$$$$$$$$$$$$$$$$$$$$$$$$$$$$$$$$     %%%%%%%

%\section*{Acknowledgments}
%The authors thank the anonymous Referees for their insightful comments, which improved the manuscript in its present form. 

%%%%%%%%%%%%%%%%%%%%%%%
%\section*{Acknowledgments} The authors thank all the three Referees for their insightful comments, which improved the manuscript in its present form.
%%%%%%%%%%%%%%%%%%%%%%%%%%%%% 
%\section*{Author declarations}
%\subsection*{Declaration of competing interest}
%The authors declare that they have no known competing financial interests or personal relationships that could have appeared to influence the work reported in this paper.
%The authors have no conflicts to disclose.
%\subsection*{Author Contributions}
%\textbf{T. D. Kaladze:} Writing--original draft, Validation, Methodology, Investigation, Formal analysis,  Conceptualization. \textbf{A. P. Misra:} Writing--review \& editing, Validation, Software, Methodology, Investigation, Conceptualization. 
%%%%%%%%%%%%%%%%%%%%%%%%%%
\subsection*{Data availability}
%The data that support the findings of this study are available from the corresponding author upon reasonable request.
%All data that support the findings of this study are included within the article (and any supplementary files).
Data will be made available on request.
%%%%%%%%%%%%%%%%%%%%%%%%%5
%\bibliographystyle{apsrev4-1} 
\bibliographystyle{cas-model2-names}
\bibliography{Reference}

\begin{thebibliography}{15}
\expandafter\ifx\csname natexlab\endcsname\relax\def\natexlab#1{#1}\fi
\providecommand{\url}[1]{\texttt{#1}}
\providecommand{\href}[2]{#2}
\providecommand{\path}[1]{#1}
\providecommand{\DOIprefix}{doi:}
\providecommand{\ArXivprefix}{arXiv:}
\providecommand{\URLprefix}{URL: }
\providecommand{\Pubmedprefix}{pmid:}
\providecommand{\doi}[1]{\href{http://dx.doi.org/#1}{\path{#1}}}
\providecommand{\Pubmed}[1]{\href{pmid:#1}{\path{#1}}}
\providecommand{\bibinfo}[2]{#2}
\ifx\xfnm\relax \def\xfnm[#1]{\unskip,\space#1}\fi
%Type = Article
\bibitem[{Abdikian et~al.(2022)Abdikian, Eghbali and Misra}]{abdikian2022}
\bibinfo{author}{Abdikian, A.}, \bibinfo{author}{Eghbali, M.},
  \bibinfo{author}{Misra, A.P.}, \bibinfo{year}{2022}.
\newblock \bibinfo{title}{Drift ion-acoustic waves in a nonuniform rotating
  magnetoplasma with two-temperature superthermal electrons}.
\newblock \bibinfo{journal}{Physica Scripta} \bibinfo{volume}{97},
  \bibinfo{pages}{045603}.
\newblock \URLprefix \url{https://dx.doi.org/10.1088/1402-4896/ac57df},
  \DOIprefix\doi{10.1088/1402-4896/ac57df}.
%Type = Article
\bibitem[{Bihlo(2008)}]{bihlo2008}
\bibinfo{author}{Bihlo, A.}, \bibinfo{year}{2008}.
\newblock \bibinfo{title}{Rayleigh–bénard convection as a nambu-metriplectic
  problem}.
\newblock \bibinfo{journal}{Journal of Physics A: Mathematical and Theoretical}
  \bibinfo{volume}{41}, \bibinfo{pages}{292001}.
\newblock \URLprefix \url{https://dx.doi.org/10.1088/1751-8113/41/29/292001},
  \DOIprefix\doi{10.1088/1751-8113/41/29/292001}.
%Type = Article
\bibitem[{Chatterjee and Misra(2021)}]{chatterjee2021}
\bibinfo{author}{Chatterjee, D.}, \bibinfo{author}{Misra, A.},
  \bibinfo{year}{2021}.
\newblock \bibinfo{title}{Effects of coriolis force on the nonlinear
  interactions of acoustic-gravity waves in the atmosphere}.
\newblock \bibinfo{journal}{Journal of Atmospheric and Solar-Terrestrial
  Physics} \bibinfo{volume}{222}, \bibinfo{pages}{105722}.
\newblock \URLprefix
  \url{https://www.sciencedirect.com/science/article/pii/S1364682621001760},
  \DOIprefix\doi{https://doi.org/10.1016/j.jastp.2021.105722}.
%Type = Article
\bibitem[{Chitre and Gokhale(1973)}]{chitre1973}
\bibinfo{author}{Chitre, S.M.}, \bibinfo{author}{Gokhale, M.H.},
  \bibinfo{year}{1973}.
\newblock \bibinfo{title}{Convective instability in a compressible atmosphere}.
\newblock \bibinfo{journal}{Solar Physics} \bibinfo{volume}{30},
  \bibinfo{pages}{309--318}.
\newblock \URLprefix \url{https://doi.org/10.1007/BF00152662},
  \DOIprefix\doi{10.1007/BF00152662}.
%Type = Article
\bibitem[{Kaladze and Misra(2023)}]{kaladze2023}
\bibinfo{author}{Kaladze, T.}, \bibinfo{author}{Misra, A.},
  \bibinfo{year}{2023}.
\newblock \bibinfo{title}{Thermal expansion of atmosphere and stability of
  vertically stratified fluids}.
\newblock \bibinfo{journal}{Physics Letters A} \bibinfo{volume}{480},
  \bibinfo{pages}{128990}.
\newblock \URLprefix
  \url{https://www.sciencedirect.com/science/article/pii/S0375960123003705},
  \DOIprefix\doi{https://doi.org/10.1016/j.physleta.2023.128990}.
%Type = Article
\bibitem[{Kaladze and Misra(2024)}]{kaladze2024}
\bibinfo{author}{Kaladze, T.D.}, \bibinfo{author}{Misra, A.P.},
  \bibinfo{year}{2024}.
\newblock \bibinfo{title}{Influence of temperature-dependent density
  inhomogeneity on the stability of atmospheric stratified fluids}.
\newblock \bibinfo{journal}{Physica Scripta} \bibinfo{volume}{99},
  \bibinfo{pages}{085013}.
\newblock \URLprefix \url{https://dx.doi.org/10.1088/1402-4896/ad5ccc},
  \DOIprefix\doi{10.1088/1402-4896/ad5ccc}.
%Type = Article
\bibitem[{Kameyama and Kinoshita(2013)}]{kameyama2013}
\bibinfo{author}{Kameyama, M.}, \bibinfo{author}{Kinoshita, Y.},
  \bibinfo{year}{2013}.
\newblock \bibinfo{title}{{On the stability of thermal stratification of highly
  compressible fluids with depth-dependent physical properties: implications
  for the mantle convection of super-Earths}}.
\newblock \bibinfo{journal}{Geophysical Journal International}
  \bibinfo{volume}{195}, \bibinfo{pages}{1443--1454}.
\newblock \URLprefix \url{https://doi.org/10.1093/gji/ggt321},
  \DOIprefix\doi{10.1093/gji/ggt321}.
%Type = Article
\bibitem[{Kopp et~al.(2021)Kopp, Tur and Yanovsky}]{kopp2021}
\bibinfo{author}{Kopp, M.}, \bibinfo{author}{Tur, A.},
  \bibinfo{author}{Yanovsky, V.}, \bibinfo{year}{2021}.
\newblock \bibinfo{title}{Nonlinear vortex structures driven by small-scale
  non-helical forces in obliquely rotating stratified fluids}.
\newblock \bibinfo{journal}{Ukrainian Journal of Physics} \bibinfo{volume}{66},
  \bibinfo{pages}{478}.
\newblock \URLprefix
  \url{https://ujp.bitp.kiev.ua/index.php/ujp/article/view/2020099},
  \DOIprefix\doi{10.15407/ujpe66.6.478}.
%Type = Book
\bibitem[{Satoh(2004)}]{satoh2004}
\bibinfo{author}{Satoh, M.}, \bibinfo{year}{2004}.
\newblock \bibinfo{title}{Atmospheric Circulation Dynamics and General
  Circulation Models}.
\newblock \bibinfo{publisher}{Springer Berlin}.
\newblock \URLprefix \url{https://doi.org/10.1007/978-3-642-13574-3}.
%Type = Article
\bibitem[{Shmerlin and Kalashnik(2013)}]{shmerlin2013}
\bibinfo{author}{Shmerlin, B.Y.}, \bibinfo{author}{Kalashnik, M.V.},
  \bibinfo{year}{2013}.
\newblock \bibinfo{title}{Rayleigh convective instability in the presence of
  phase transitions of water vapor. the formation of large-scale eddies and
  cloud structures}.
\newblock \bibinfo{journal}{Physics-Uspekhi} \bibinfo{volume}{56},
  \bibinfo{pages}{473}.
\newblock \URLprefix \url{https://dx.doi.org/10.3367/UFNe.0183.201305d.0497},
  \DOIprefix\doi{10.3367/UFNe.0183.201305d.0497}.
%Type = Article
\bibitem[{Turi and Misra(2022)}]{turi2022}
\bibinfo{author}{Turi, J.}, \bibinfo{author}{Misra, A.P.},
  \bibinfo{year}{2022}.
\newblock \bibinfo{title}{Magnetohydrodynamic instabilities in a
  self-gravitating rotating cosmic plasma}.
\newblock \bibinfo{journal}{Physica Scripta} \bibinfo{volume}{97},
  \bibinfo{pages}{125603}.
\newblock \URLprefix \url{https://dx.doi.org/10.1088/1402-4896/ac9ca6},
  \DOIprefix\doi{10.1088/1402-4896/ac9ca6}.
%Type = Article
\bibitem[{Turkyilmazoglu and Duraihem(2024)}]{turky2024}
\bibinfo{author}{Turkyilmazoglu, M.}, \bibinfo{author}{Duraihem, F.Z.},
  \bibinfo{year}{2024}.
\newblock \bibinfo{title}{Generalized mean state of the isothermal
  darcy–benard problem and its instability onset}.
\newblock \bibinfo{journal}{European Journal of Mechanics - B/Fluids}
  \bibinfo{volume}{103}, \bibinfo{pages}{334--342}.
\newblock \URLprefix
  \url{https://www.sciencedirect.com/science/article/pii/S0997754623001565},
  \DOIprefix\doi{https://doi.org/10.1016/j.euromechflu.2023.11.002}.
%Type = Article
\bibitem[{Turkyilmazoglu and Siddiqui(2023a)}]{turky2023a}
\bibinfo{author}{Turkyilmazoglu, M.}, \bibinfo{author}{Siddiqui, A.A.},
  \bibinfo{year}{2023}a.
\newblock \bibinfo{title}{The instability onset of generalized isoflux mean
  flow using brinkman-darcy-bénard model in a fluid saturated porous channel}.
\newblock \bibinfo{journal}{International Journal of Thermal Sciences}
  \bibinfo{volume}{188}, \bibinfo{pages}{108249}.
\newblock \URLprefix
  \url{https://www.sciencedirect.com/science/article/pii/S1290072923001102},
  \DOIprefix\doi{https://doi.org/10.1016/j.ijthermalsci.2023.108249}.
%Type = Article
\bibitem[{Turkyilmazoglu and Siddiqui(2023b)}]{turky2023b}
\bibinfo{author}{Turkyilmazoglu, M.}, \bibinfo{author}{Siddiqui, A.A.},
  \bibinfo{year}{2023}b.
\newblock \bibinfo{title}{The instability onset of generalized isoflux mean
  flow using brinkman-darcy-bénard model in a fluid saturated porous channel}.
\newblock \bibinfo{journal}{International Journal of Thermal Sciences}
  \bibinfo{volume}{188}, \bibinfo{pages}{108249}.
\newblock \URLprefix
  \url{https://www.sciencedirect.com/science/article/pii/S1290072923001102},
  \DOIprefix\doi{https://doi.org/10.1016/j.ijthermalsci.2023.108249}.
%Type = Article
\bibitem[{Zhang et~al.(2024)Zhang, Liu, Wu and Li}]{zhang2024}
\bibinfo{author}{Zhang, F.}, \bibinfo{author}{Liu, W.}, \bibinfo{author}{Wu,
  L.}, \bibinfo{author}{Li, J.}, \bibinfo{year}{2024}.
\newblock \bibinfo{title}{Chaotic model and control of an atmospheric
  convective system coupled with large-scale circulation}.
\newblock \bibinfo{journal}{Physica Scripta} \bibinfo{volume}{99},
  \bibinfo{pages}{045213}.
\newblock \URLprefix \url{https://dx.doi.org/10.1088/1402-4896/ad2bc1},
  \DOIprefix\doi{10.1088/1402-4896/ad2bc1}.

\end{thebibliography}
\nopagebreak
\end{document}